\newif\ifREVTEXIV
\REVTEXIVtrue

\ifREVTEXIV
\documentclass[prd,aps,floatfix]{revtex4}
\usepackage{graphicx}
\else
\documentstyle[prd,aps,epsf]{revtex}
\fi

\newcommand{\p}{\partial}
\newcommand{\kslash}{k\kern-1ex /}
\newcommand{\pslash}{p\kern-1ex /}
\newcommand{\qslash}{q\kern-1ex /}
\newcommand{\lslash}{l\kern-1ex /}
\newcommand{\sslash}{s\kern-1ex /}
\newcommand{\Dslash}{{\cal D}\kern-1.5ex /}

\newcommand{\beqa}{\begin{eqnarray}}
\newcommand{\eeqa}{\end{eqnarray}}

\newcommand{\be}{\begin{equation}}
\newcommand{\ee}{\end{equation}}
\newcommand{\ben}{\begin{eqnarray}}
\newcommand{\een}{\end{eqnarray}}
\newcommand{\nn}{\nonumber}
\def\lsim{\raise0.3ex\hbox{$<$\kern-0.75em\raise-1.1ex\hbox{$\sim$}}}
\def\gsim{\raise0.3ex\hbox{$>$\kern-0.75em\raise-1.1ex\hbox{$\sim$}}}
\def\simgt{\rlap{\lower 3.5 pt\hbox{$\mathchar
\sim$}}\raise 1pt \hbox {$>$}}
\def\simlt{\rlap{\lower 3.5 pt\hbox{$\mathchar
\sim$}}\raise 1pt \hbox {$<$}}

\newcommand{\latt}{{\rm latt}}

\newcommand{\mf}{{\rm MF}}
\newcommand{\msbar}{{\overline {\rm MS}}}

\newcommand{\ce}{{\it c}_{\it E}}
\newcommand{\cb}{{\it c}_{\it B}}
\newcommand{\lqcd}{{\Lambda}_{\rm QCD}}

\newcommand{\lo}{{(0)}}
\newcommand{\nlo}{{(1)}}
\newcommand{\mplo}{{m_p^{(0)}}}

\newcommand{\mplomf}{{{\tilde m_p}^{(0)}}}

\newcommand{\mph}{{m_{p1}}}
\newcommand{\mpl}{{m_{p2}}}
\newcommand{\mpllo}{{m_{p2}^{(0)}}}
\newcommand{\mphlo}{{m_{p1}^{(0)}}}
\newcommand{\mpllomf}{{{\tilde m_{p2}}^{(0)}}}
\newcommand{\mphlomf}{{{\tilde m_{p1}}^{(0)}}}
\newcommand{\mqlomf}{{{\tilde m_{q}}^{(0)}}}
\newcommand{\mQlomf}{{{\tilde m_{Q}}^{(0)}}}

\def\Ovec{\Delta\hspace{-0.3cm}\raisebox{1.8ex}{$\rightarrow$}}
\def\Antivec{\Delta\hspace{-0.3cm}\raisebox{1.8ex}{$\leftarrow$}}

\ifREVTEXIV
\newdimen\epsfxsize
\newdimen\epsfysize

\def\epsfbox#1{\includegraphics[width=\epsfxsize]{#1}}
\renewcommand{\address}[1]{\affiliation{#1}}

\begin{document}
\fi

\draft
\title{
\vspace*{-2.cm}
\begin{flushright}
{\normalsize UTHEP-537}\\
\end{flushright}
First Nonperturbative Test of a Relativistic Heavy
Quark Action\\ in Quenched Lattice QCD
}
\author{
    Y.~Kayaba\rlap,$^{\rm 1}$
        S.~Aoki\rlap,$^{\rm 1,2}$
    M.~Fukugita\rlap,$^{\rm 3}$
     Y.~Iwasaki\rlap,$^{\rm 1}$
      K.~Kanaya\rlap,$^{\rm 1}$
   Y.~Kuramashi\rlap,$^{\rm 1,4}$
       M.~Okawa\rlap,$^{\rm 5}$
            A.~Ukawa,$^{\rm 1,4}$  and
          T.~Yoshi\'e$^{\rm 1,4}$\\
(CP-PACS Collaboration)
}
\address{
$^{\rm 1}$
Graduate School of Pure and Applied Sciences, University
of Tsukuba, Tsukuba 305-8571, Japan\\
$^{\rm 2}$
Riken BNL Research Center, Brookhaven National Laboratory Upton, NY11973, USA \\
$^{\rm 3}$
Institute for Cosmic Ray Research,
University of Tokyo, 
Kashiwa, Chiba 277-8582, Japan \\
$^{\rm 4}$
Center for Computational Sciences, University of
Tsukuba, Tsukuba 305-8577, Japan\\
$^{\rm 5}$
Department of Physics,
Hiroshima University, 
Higashi-Hiroshima, Hiroshima 739-8526, Japan \\
}
\date{\today}

\ifREVTEXIV
\else
  \begin{document}
  \draft
  \tightenlines
  \maketitle
\fi


\begin{abstract}

We perform a numerical test of a relativistic heavy quark(RHQ) action, 
recently proposed by Tsukuba group, 
in quenched lattice QCD at $a\simeq 0.1$ fm.
With the use of the improvement parameters previously 
determined at one-loop level for the RHQ action,
we investigate a restoration of rotational symmetry for heavy-heavy and 
heavy-light meson systems around the charm quark mass. 
We focused on two quantities,
the meson dispersion relation and the pseudo-scalar meson decay constants.
It is shown that the RHQ action significantly reduces 
the discretization errors due to the 
charm quark mass.
We also calculate the S-state hyperfine splittings 
for the charmonium and charmed-strange mesons
and the $D_s$ meson decay constant. 
The remaining discretization errors in the physical quantities
are discussed.

\end{abstract}

\pacs{PACS number(s): }

\ifREVTEXIV
\maketitle
\fi


\section{Introduction}

For a search of new physics beyond the standard model through flavor
physics, a precise determination of physical
quantities such as quark masses and hadronic matrix elements associated
with heavy mesons is required within a few \% accuracy.
Although, in principle, lattice QCD calculation is an ideal tool 
for this purpose, it suffers from large discretization effects due to
the charm and bottom quark masses:
$a m_c \simgt 0.3$, $a m_b \simgt 1$ in quenched approximation and 
$a m_c \simgt 0.5$, $a m_b \simgt 1.5$ in unquenched simulation
with current computational resources.
If we adopt the $O(a)$ improved Wilson quark action for the heavy quarks, 
the leading cutoff errors are expected to be
$O((a m_Q)^n )$.
In order to achieve a few $\%$ accuracy,
it is necessary to reduce the discretization errors for the heavy quarks 
to the same level for the light quarks,
which is $O((a \Lambda_{\rm QCD})^2)$.

For this end an on-shell $O(a \lqcd)$ improved
RHQ action has been proposed in Ref.\cite{akt}, which
extends the well known on-shell improvement
program\cite{sym1,sym2,sw,alpha,onshell}
to massive quarks with $a m_Q \sim O(1)$.
The action works better as the lattice spacing becomes smaller.
This is a fascinating feature from a view point of 
controlling the systematic errors coming from the cutoff effects.
Another important point is that the action allows to treat  
the charm and bottom quarks simultaneously.
In case of NRQCD, widely used for calculation of
the bottom quark physics, on the other hand, it is theoretically impossible to take 
the continuum limit and difficult to treat the charm quark.

The explicit form of the RHQ action is given by
\ben
S^{\rm RHQ} &=& \sum_x\left[ m_0{\bar q}(x)q(x)
   +{\bar q}(x)\gamma_4 D_4q(x)
   +\nu \sum_i {\bar q}(x)\gamma_i D_i q(x)\right. \nn\\
&& \left.-\frac{r_t a}{2} {\bar q}(x)D_4^2 q(x)
   -\frac{r_s a}{2} \sum_i {\bar q}(x)D_i^2 q(x)\right. \label{eq:action_q}\\
&& \left.-\frac{ig a}{2}\ce \sum_i {\bar q}(x)\sigma_{4i}F_{4i} q(x)
   -\frac{ig a}{4}\cb \sum_{i,j} {\bar q}(x)\sigma_{ij}F_{ij} q(x)
   \right], \nn
\een
where four improvement parameters
$\nu$, $r_s$, $c_E$ and $c_B$ are relevant, while $r_t$ is redundant.
The leading cutoff effects of $O((a m_Q)^n)$ in this formulation
can be removed by adjusting $\nu$ and quark field
renormalization factor as a function of $a m_Q$ and
gauge coupling constant $g$.
We can also remove the next-to-leading cutoff effects 
of $O((am_Q)^n a\lqcd)$
by adjusting $r_s$, $c_E$ and $c_B$. 
Note that it is recently pointed out that one can remove 
$O((am_Q)^n a\lqcd)$ errors in the spectral quantities such as masses by 
adjusting only two parameters, $c_B=c_E$ and $\nu$\cite{Christ}. 
However it is still true that 4 parameters, $\nu$, $r_s$, $c_E$ and
$c_B$, are necessary to remove all $O((am_Q)^n a\lqcd)$ errors in  
on-shell matrix elements.
Once these are achieved, we are left with
only the cutoff effects of $O(f_2(a m_Q) (a \lqcd)^2)$
where $f_2(a m_Q)$ is an analytic function of $a m_Q$ around $a m_Q=0$.
We assume $f_2(a m_Q) \sim O(1)$ for the massive quarks with $m_Q a \sim O(1)$.
In the massless limit $\nu$ and  $r_s$ become unity and $c_E$ and $c_B$ 
agree with $c_{\rm SW}$.
Since, at present, it is difficult to determine $r_s$, $c_E$, $c_B$ 
nonperturbatively,  we have
performed a perturbative determination of the four
parameters at one-loop level in a mass dependent
way\cite{cEcB}.
In this case  
the remaining leading cutoff effects are  $O(\alpha_s^2 f_0^{(2)}(a m_Q))$
for the RHQ action.
Similarly, we have also determined the 
mass-dependent renormalization constants and the improvement coefficients
for the vector and axial-vector currents at one-loop level\cite{zAcA}.

In this work we study the discretization effects of the RHQ
action with the perturbatively determined improvement parameters
using two different gauge actions 
in quenched approximation at a finite lattice spacing of $a \simeq 0.1$fm.
We employ four heavy quark masses around the charm quark mass.
In order to investigate the discretization effects, we calculate the
dispersion relation both for the heavy-heavy and heavy-light mesons
and the space-time symmetry of the pseudo-scalar meson decay matrix elements.
For comparison, the same calculation is repeated with the heavy clover 
quark action. We observe sufficient reduction of
$(am_Q)^n$ errors in the RHQ action.
In addition 
we extract the charmed-strange and charmonium hyperfine splittings 
and the $D_s$ meson decay constant $f_{D_s}$ at the physical charm quark mass.
We compare our results with previous ones and 
discuss the cutoff effects on these quantities 
with the RHQ quark action.  

This paper is organized as follows. 
In Sect.~\ref{Sec:FORMULATION} we explain the RHQ action 
and fix the notations. Simulation details are summarized in
Sect.~\ref{Sec:Simulation details}. In Sect.~\ref{Sec:Results}
we give a comparison between the RHQ action and the clover action
by showing the dispersion relation and 
the space-time symmetry.
We also present the results of physical quantities 
such as the hyperfine splittings and the decay constants,
and discuss their cutoff errors in detail.
A comparison of our results with the previous ones for
the hyperfine splittings and the decay constants is also shown.
In Sect.~\ref{Sec:Conclusions} we give our conclusion.
A brief review of recent works related to this
formulation can be found in Ref.\cite{KrmshNote}.

\section{FORMULATION}
\label{Sec:FORMULATION}

\subsection{Actions}

For the gauge part we employ a renormalization-group (RG)
improved gauge action proposed by Iwasaki\cite{iwasaki} 
as well as the ordinary plaquette gauge action. 
For the quark part we use the clover quark action\cite{sw}
for the light quarks and 
the RHQ action for the heavy quarks.
We rewrite the RHQ action of Eq.(\ref{eq:action_q})
to the following form with the use of hopping parameter $\kappa$, 
which is more suitable for numerical simulations:
\beqa
S^{\rm RHQ} & = & \sum_{x,y}\overline{q}(x) D_{x,y}q(y),\\
D_{x,y} & = & \delta_{xy}
- \kappa \sum_{k=1,3} \left\{(r_s-\nu \gamma_k)U_{x,k}
\delta_{x+\hat{k},y}
      + (r_s+\nu \gamma_k)U_{x,k}^{\dag}
\delta_{x,y+\hat{k}} \right\} \nn\\
& & -\kappa \left\{ (r_t-\gamma_4)U_{x,4}
\delta_{x+\hat{4},y}
      + (r_t+\gamma_4)U_{x,4}^{\dag}
\delta_{x,y+\hat{4}}\right\} \nn\\
& & - \, \delta_{xy} c_B \kappa \sum_{i < j}
         \sigma_{ij} F_{ij}(x)
- \, \delta_{xy} c_E \kappa \sum_{i}
         \sigma_{4i} F_{4i}(x) ,
\label{eq:RHQ action}
\eeqa
where the field strength $F_{\mu\nu}$ in the clover terms
is expressed as
\ben 
F_{\mu\nu}(x)&=&\frac{1}{8}\sum_{i=1}^{4}
\left(U_i(x)-U_i^\dagger(x)\right), \\
U_1(x)&=&U_{x,\mu}U_{x+{\hat \mu},\nu}
         U^\dagger_{x+{\hat \nu},\mu}U^\dagger_{x,\nu}, \\
U_2(x)&=&U_{x,\nu}U^\dagger_{x-{\hat \mu}+{\hat \nu},\mu}
         U^\dagger_{x-{\hat \mu},\nu}U_{x-{\hat \mu},\mu}, \\
U_3(x)&=&U^\dagger_{x-{\hat \mu},\mu}U^\dagger_{x-{\hat \mu}-{\hat \nu},\nu}
         U_{x-{\hat \mu}-{\hat \nu},\mu}U_{x-{\hat \nu},\nu}, \\
U_4(x)&=&U^\dagger_{x-{\hat \nu},\nu}U_{x-{\hat \nu},\mu}
         U_{x+{\hat \mu}-{\hat \nu},\nu}U^\dagger_{x,\mu}.
\een 
As mentioned in the introduction,
the parameters $\nu$, $r_s$, $c_B$ and $c_E$ 
are already determined as a function of the heavy quark mass up to one-loop
level. In this work we choose $r_t=1$.

\subsection{Improvement of axial vector current}

The form of renormalized axial vector current 
with the $O(a \lqcd)$ improvement is given in Ref.\cite{zAcA}:
\beqa
A^{R}_\mu(x) &=&
\sqrt{2 \kappa_q} \sqrt{2 \kappa_Q} Z_{A_\mu} \left[~
{\bar q(x)} \gamma_\mu\gamma_5 Q(x)
-c_{A_\mu}^+  \{{\bar q(x)} \Delta_\mu^+ \gamma_5 Q(x)\}
-c_{A_\mu}^-  \{{\bar q(x)}\Delta_\mu^- \gamma_5 Q(x)\}
\right. \nn \\
&&\left.
+c_{A_\mu}^L \{{\vec \Delta_i}{\bar q(x)}\} \gamma_i
\gamma_\mu\gamma_5 Q(x) 
-c_{A_\mu}^R {\bar q(x)}\gamma_\mu\gamma_5 \gamma_i
\{{\vec \Delta_i} Q( x)\}
 ~\right] ,
\label{eq:RHQcrrntImpr} 
\eeqa
where $q$ and $Q$ denote the light and heavy quark fields, respectively.
$Z_{A_\mu}$ is the finite renormalization factor connecting
the lattice to the continuum $\msbar$ scheme ($Z_A =
Z_{A_\mu}^{\rm latt}/Z_{A_\mu}^{\rm cont}$) as defined in
Ref.\cite{zAcA}.
Improvement coefficients for the temporal direction are in general
different from those for the spatial direction: 
$c_{A_4}^{\pm,R,L}\not=c_{A_k}^{\pm,R,L}$.
An additional overall factor $\sqrt{2 \kappa_q} \sqrt{2 \kappa_Q}$
is associated with the field redefinition 
in the RHQ action of Eq.(\ref{eq:RHQ action}).
$Z_{A_\mu}$ and $c_{A_\mu}^{\{\pm,R,L\}}$
are calculated as a function of the quark masses $a m_Q$ and
$a m_q$ up to one-loop level\cite{zAcA}.
With the aid of the equation of motion we can always choose
$c^R_{A_4}=c^L_{A_4}=0$.
Covariant lattice derivatives 
$\Delta_\mu^+$ and $\Delta_\mu^-$ are defined as 
$\Delta_\mu^+ = \Ovec_\mu + \Antivec_\mu$ and 
$\Delta_\mu^- = \Ovec_\mu - \Antivec_\mu$, where $\Ovec_\mu$ and
$\Antivec_\mu$ are lattice derivative acting on the right or
left field as
\beqa
{\bar q(x)}~\gamma_5~\Ovec_\mu~ Q(x) 
&=& {\bar q(x)}~\gamma_5~\frac{1}{2}[~U_\mu(x) Q(x+{\hat
\mu}) 
-U_\mu^\dagger(x-{\hat \mu}) Q(x-{\hat \mu})~], 
\label{Eq:lattDrvtvR}\\
{\bar q(x)}~\Antivec_\mu~\gamma_5~Q(x) &=&\frac{1}{2}[~{\bar
q(x+{\hat \mu})}
U_\mu^\dagger(x) -{\bar q(x-{\hat
\mu})}U_\mu(x-{\hat\mu})~]
~\gamma_5~Q(x).
\label{Eq:lattDrvtvL}
\eeqa
Note that the definition of the improved
current of Eq.(\ref{eq:RHQcrrntImpr}) is slightly modified from that
in Ref.\cite{zAcA}. The difference will be 
explained in Appendix A.
Hereafter we use a short-handed notation for the improvement terms such as
\beqa
O_\mu^+ &\equiv& -c_{A_\mu}^+ \{{\bar q(x)} \Delta_\mu^+
\gamma_5 Q(x)\},
\label{eq:A corr term +} \\
O_\mu^- &\equiv& -c_{A_\mu}^- \{{\bar q(x)} \Delta_\mu^-
\gamma_5 Q(x)\}, \\
O_k^L &\equiv& +c_{A_k}^L \{{\Ovec_i}{\bar
q(x)}\} \gamma_i \gamma_k\gamma_5 Q(x), \\
O_k^R &\equiv& -c_{A_k}^R {\bar q(x)}
\gamma_k\gamma_5 \gamma_i \{{\Ovec_i} Q(x)\} .
\label{eq:A corr term R}
\eeqa

The pseudo-scalar meson decay constant is 
defined as 
\beqa
\left< 0| A_\mu^R(0) |PS(p) \right> &=& i p_\mu f_{PS}
(A_\mu),
\label{Eq:fPS:def}
\eeqa
where $|PS(p)\rangle$ is the pseudo-scalar meson state with momentum
$p$.
The decay constant denoted
as $f_{PS} (A_\mu)$ identifies which component of the
axial vector current is used.
In the continuum limit $f_{PS} (A_\mu)$ for different $\mu$ should agree with
each other.

\section{simulation details}
\label{Sec:Simulation details}

\subsection{Simulation parameters}

We employ a single value of the gauge
coupling constant, $\beta=2.6$ for the Iwasaki action and
$\beta=6.0$ for the plaquette action, 
on a $L^3\times T=24^3\times 48$ lattice.
Gauge configurations are generated by a 5-hit pseudo
heat bath update supplemented by four over-relaxation steps.
These configurations are then fixed to the
Coulomb gauge at every 100 sweeps for the Iwasaki gauge
action and 200 sweeps for the plaquette gauge action. 
We have accumulated $300$ gauge configurations for each gauge
action with the RHQ action and 250 with the heavy clover quark action.
With the use of the Sommer scale $r_0=0.5$fm the lattice cutoffs are 
determined as $a^{-1}(r_0)=2.0129(46) [{\rm GeV}]$ \cite{r0Necco}
and $a^{-1}(r_0)=2.1184(94) [{\rm GeV}]$
\cite{r0Sommer}, respectively. 
The spatial lattice size in physical unit is approximately $2.4$fm, 
which is large enough for the charmed mesons.

Simulation parameters for the quark part are summarized in
Table~\ref{Tab:Sim:Iwasaki+RHQ} for the Iwasaki 
and the plaquette gauge actions, where 
$M_{PS}/M_V$ represents the 
pseudo-scalar to vector meson mass ratio 
of the light-light and heavy-heavy mesons.
For each gauge action we adopt three values of the light quark masses
corresponding to  $M_{PS}/M_V\simeq 0.55-0.78$ 
to cover the strange quark mass and four values of the heavy quark masses to
sandwich the charm quark mass.

For the light quarks we use the clover quark action with
the nonperturbative value of the clover coefficient:
$c_{\rm SW}^{\rm NP}=1.50(5)$\cite{NPCL.CPPACS} for the Iwasaki action and
$1.769$\cite{NPCL.ALPHA} for the plaquette action. 
Here it is noted that $c_{\rm SW}^{\rm NP}=1.50(5)$ for 
the Iwasaki action is taken from
the preliminary result obtained in the infinite volume limit,
which is 6 \% larger than the final value $c_{\rm SW}^{\rm NP}=1.41$ 
of Ref.\cite{NPCL.CPPACS} defined on a fixed physical volume.
For the heavy quarks we adopt the RHQ action with the improvement
parameters $\nu$,  $r_s$, $c_E$ and $c_B$
determined up to one-loop level with the mean-field(MF) 
improvement,
details of which are explained in Appendix A.
In order to remove $O(a \lqcd)$
errors  at the massless point,
we replace a massless part of $c_E$ and $c_B$
by their nonperturbative value $c_{\rm SW}^{\rm NP}$ as
\beqa
c_{E,B} = c_{E,B}^{\rm PT}(a m_{\rm pole})
-c_{E,B}^{\rm PT}(a m_{\rm pole}=0)
+ c_{\rm SW}^{\rm NP} ,
\label{Eq:cEcBNPshift:Nf0}
\eeqa
where the superscript PT represents the perturbative value up to
one-loop level.

For the light-light current we use
nonperturbative values of the renormalization factor 
and the improvement coefficients for the plaquette gauge action: 
$Z_A^{\rm NP}=0.807$, $b_A^{\rm NP}=1.28$ and 
$c_A^{\rm NP}=0.037$\cite{NPzAcA:LALM:3}.
For the Iwasaki gauge action
we employ the mean-field improved values: $Z_A=0.86057$, $b_A=1.19998$ 
and $c_A=0.00864$\cite{gmass}. 
At present nonperturbative values are not available for this action.
For the heavy-light and heavy-heavy currents, on the other hand, we use
the mean-field improved values for $c_{A_\mu}^{\{\pm,L,R\}}$ 
and $Z_{A_\mu}$ at the one-loop level (see Appendix A) for both gauge actions. 
In case of the plaquette gauge action we replace 
the massless part of $c_{A_\mu}^+$ and $Z_{A_\mu}$ by 
the nonperturbative ones,
$c_A^{\rm NP}$ and $Z_A^{\rm NP}$:
\beqa
c_{A_\mu}^+ &=&
g^2 c_{A_\mu}^{+,{\rm PT}}(a m_{\rm pole})
- g^2 c_{A_\mu}^{+,{\rm PT}}(a m_{\rm pole}=0)
+ c_A^{+,{\rm NP}}, \label{Eq:cANPshift}\\
Z_{A_\mu} &=&
Z_{A_\mu}^{\rm PT}(a m_{\rm pole})
-Z_{A_\mu}^{\rm PT}(a m_{\rm pole}=0)
+ Z_{A_\mu}^{\rm NP} .
\label{Eq:zANPshift}
\eeqa

In order to investigate a degree of improvement for the RHQ action
we have made an additional simulation
using the clover quark action both for the heavy and light quarks.
Simulation parameters are given in
Table \ref{Tab:Sim:Iwasaki+CL}, where we employ
one value of the light quark mass and
three values of the heavy quark masses roughly equal to 
lighter three for the RHQ action.

On each configuration fixed with the Coulomb gauge, we invert 
the quark matrix employing the BiCGstab algorithm
with the stopping condition that the residual must be smaller than
$1.0\times 10^{-14}$.  
For the heavy quarks we perform a fixed number of iterations.
We choose $2T=96$ such that the stopping condition is always 
satisfied and it is assured that the heavy quarks can propagate from the
origin to any point on the lattice. 
For both the light and heavy quark propagators we employ
not only a local source but also an exponentially 
smeared source with a form of 
$A \exp(-B r)$, where smearing parameters $A$ and $B$ are tuned 
to enhance an overlap with the ground state.
Numerical values of $A$ and $B$ are listed in
Table~\ref{Tab:Sim:Smear para} for each combination of the gauge and
quark actions.

\subsection{Measurement of two-point functions}
\label{sec:correlator measurements}

We measure the S-state ({\it i.e.} pseudo-scalar and vector) meson spectra
for the light-light(L-L), heavy-light(H-L) and heavy-heavy(H-H) systems
using the correlation functions projected onto zero spatial momentum state: 
\beqa
\sum_{\vec{x}} \langle {\cal O}_L(\vec{x},t) {\cal
O}_{S,L}^\dagger(0)\rangle, 
\label{Eq:crrltr:msn}
\eeqa
where ${\cal O} = P$ or $V$ is understood. 
The subscripts $S$ and $L$ represent the smeared and local
operators, respectively. We always adopt a local sink while taking 
both the local and smeared sources.
Note that both the quark and anti-quark fields in ${\cal O}_S$ are smeared.

To extract the pseudo-scalar meson decay constant for the L-L, H-L, H-H systems,
we calculate the correlation function
\beqa
\sum_{\vec{x}} \langle A_4^{\rm impr}(\vec{x},t)
P_S^\dagger(0)\rangle ,
\label{Eq:crrltr:dcy}
\eeqa
where the superscript impr represents
the $O(a)$ improved current given in
Eq.(\ref{eq:RHQcrrntImpr}).

We also measure the meson correlation functions with 
finite spatial momenta given by
\beqa
a {\vec p} = \frac{2\pi}{L} \times \{(1,0,0),\ (1,1,0)\}.
\label{Eq:latMom}
\eeqa
These correlation functions are used to
calculate the dispersion relation of the S-state mesons
and also
to extract the decay constant using the temporal and spatial components 
of the axial vector current.

\subsection{Fitting procedure}

The correlation functions in Eq.(\ref{Eq:crrltr:msn}) are expected to take the
following form for a large euclidean time separation: 
\beqa
\Sigma_{\vec{x}} \langle {\cal O}_L(\vec{x},t) {\cal
O}_{S,L}^\dagger(0)\rangle  = \frac{Z_{{\cal O}_L} Z^\dagger_{{\cal O}_{S,L}} }{ 2
a M } {\rm e}^{-a M T/2} \cosh(a M (T/2-t)), 
\label{Eq:Corr:FitA}\\
\Sigma_{\vec{x}} \langle A^{\rm impr}_\mu(\vec{x},t)
P_S^\dagger(0)\rangle =
\frac{Z_{A^{\rm impr}_\mu}Z^\dagger_{P_S} }{ 2
a M } {\rm e}^{-a M T/2} \sinh(a M (T/2-t)) ,
\label{Eq:Corr:FitB}
\eeqa
where $M$ is a mass of the ground state allowed to couple to the operator.
Matrix elements in the above expressions
are given by
\beqa
Z_{P_{S,L}} &=& \langle 0|P_{S,L}(0)|PS(\vec{p}=\vec{0})\rangle,
\\
Z_{V_{S,L}} &=& \langle 0|V_{S,L}(0)|V(\vec{p}=\vec{0})\rangle,
\\
Z_{A^{\rm impr}_\mu} &=& \langle 0|A_\mu^{\rm
impr}(0)|PS(\vec{p}=\vec{0})\rangle ,
\label{eq:zaimpr matrix el}
\eeqa
where $|PS(\vec{p}=\vec{0})\rangle$ and $|V(\vec{p}=\vec{0})\rangle$ represent the
pseudo-scalar and vector meson states at rest.
We first extract $aM$ by fitting the correlators of Eq.(\ref{Eq:Corr:FitA}), 
and then perform a fit of
Eq.(\ref{Eq:Corr:FitB}) with $a M$ fixed. We employ the same fitting
procedure for the correlation functions with finite spatial momenta.
Since our statistics are not sufficient to incorporate
correlations between different time slices,
we always use the uncorrelated fit for our analysis. 
We estimate statistical errors by the
jackknife method with a bin size of 10
configurations to eliminate autocorrelations.
In case that the correlated fit is possible, 
we use it to check the results obtained by the uncorrelated fit.
We find  
that the results are consistent within statistical errors. 

\begin{figure}
\vspace{0.8cm}
\centerline{ 
\epsfxsize=8.0cm
\epsfbox{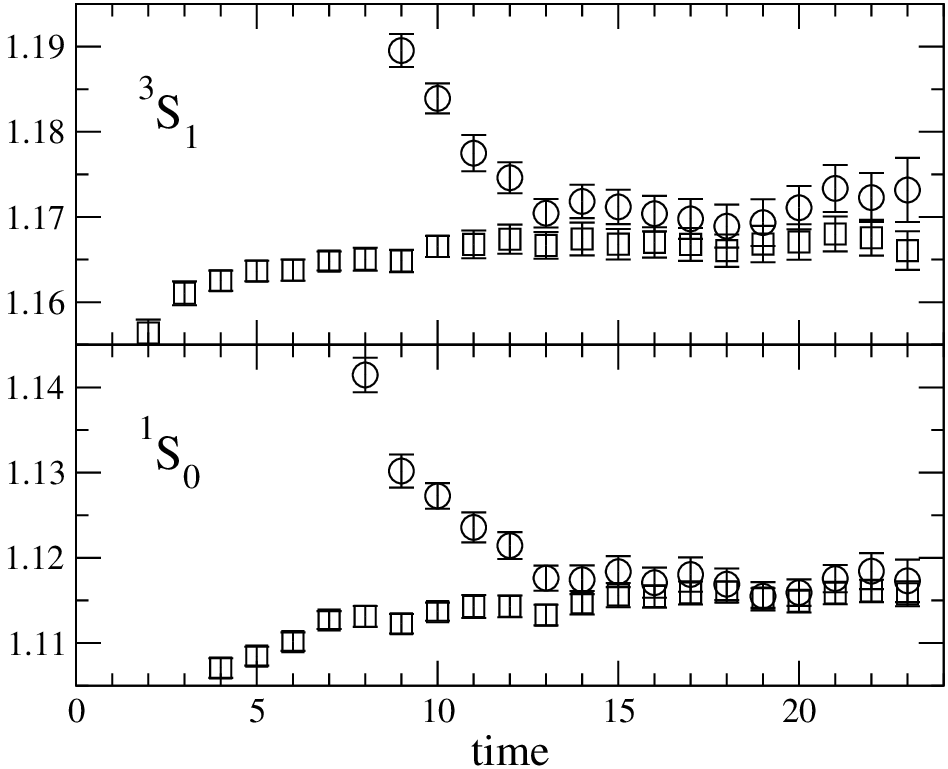}
\hspace{1cm}
\epsfxsize=8.0cm
\epsfbox{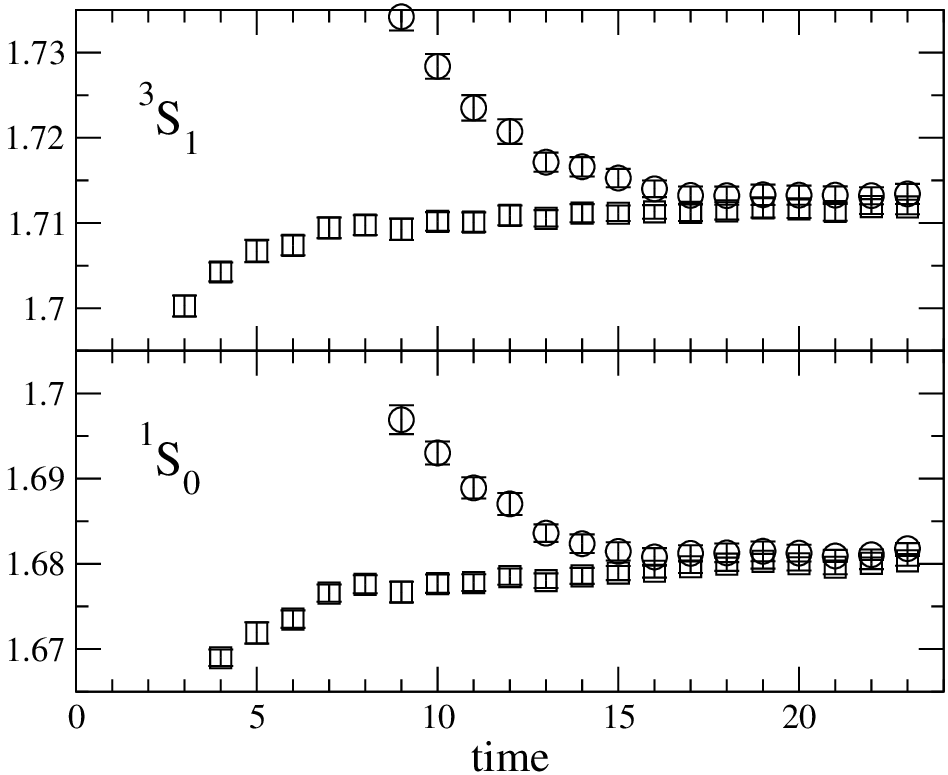}}
\caption{Effective mass plots for the H-L(left) and H-H(right)
mesons with zero spatial momentum
in case of $\kappa_{heavy}=\kappa_6$ and $\kappa_{light}=\kappa_3$
with the Iwasaki gauge action.
Circles represent the local source correlators and squares 
for the smeared source correlators.}
\label{Fig:EffMss}
\end{figure}

\begin{figure}
\centerline{ 
\epsfxsize=8.0cm
\epsfbox{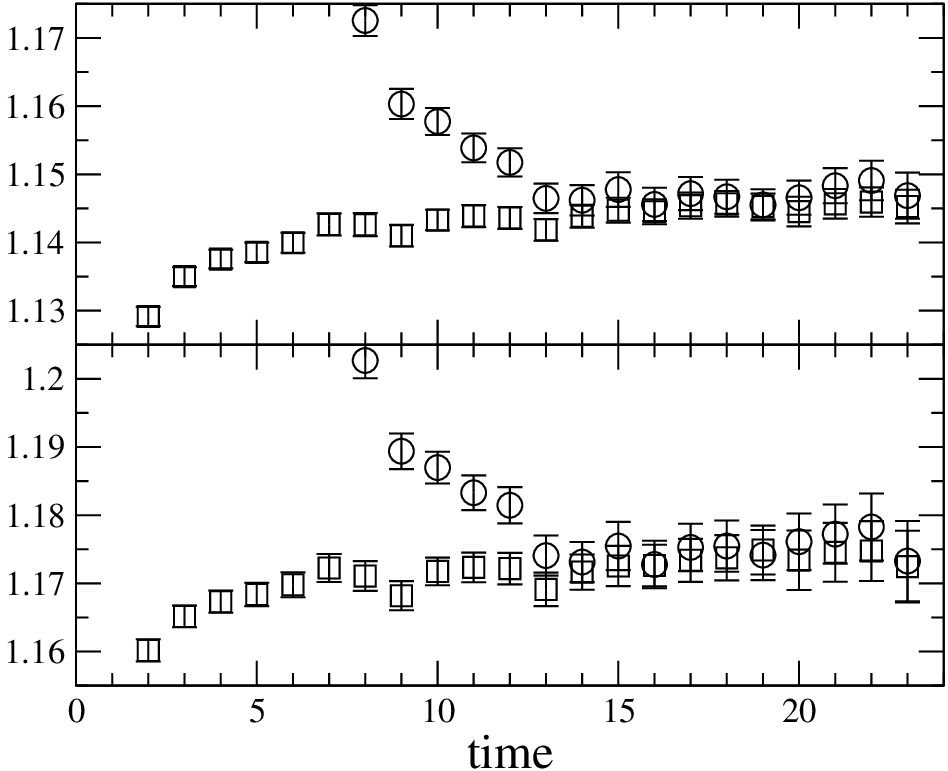}
\hspace{1cm}
\epsfxsize=8.0cm
\epsfbox{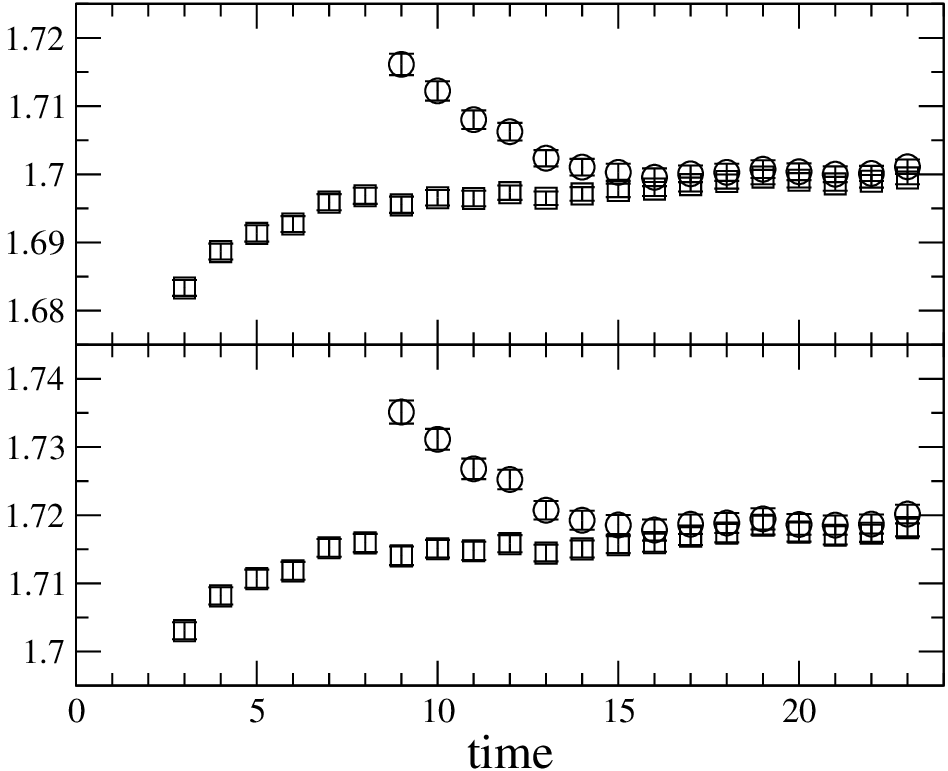}}
\caption{Effective mass plots for the H-L(left) and H-H(right)
pseudoscalar mesons with $\vert a{\vec p}\vert^2=(2\pi/L)^2$ (top) and
$\vert a{\vec p}\vert^2=2(2\pi/L)^2$ (bottom)
in case of $\kappa_{heavy}=\kappa_6$ and $\kappa_{light}=\kappa_3$ 
with the Iwasaki gauge action.
Circles represent the local source correlators and squares 
for the smeared source correlators.}
\label{Fig:EffMssP}
\end{figure}

The fitting ranges summarized
in Table~\ref{tab:fitrng} are chosen by investigating effective mass plots of 
the meson correlators presented in
Fig.\ref{Fig:EffMss}, where we take $\kappa_3$ for the light quark 
and $\kappa_6$ for the heavy quark as a representative case.
Note that $\kappa_6$ roughly corresponds to the charm quark mass.
We take similar fitting ranges for the correlators with finite spatial momenta,
which are given in Table\ref{tab:fitrng}. 
Figure~\ref{Fig:EffMssP} shows
effective mass plots for the pseudo-scalar meson
correlators with finite spatial momenta.

\section{RESULTS}
\label{Sec:Results}

\subsection{Dispersion relation and space-time
interchange symmetry}

In case that the improvement parameters are perturbatively determined
up to one-loop level,
the leading cutoff errors in the RHQ action is theoretically expected 
to be $O(\alpha_s^2 f_0^{(2)}(a
m_Q))$, where $f_0^{(2)}(a m_Q) \sim
O(1)$ is assumed for $a m_Q \sim O(1)$.
We numerically check this theoretical
expectation 
by investigating the dispersion relation of the S-state mesons and the
space-time interchange symmetry for the pseudo-scalar
meson decay constant.
These quantities are sensitive to the cutoff effects
for the heavy quarks, and hence suitable to 
estimate a size of $f_0^{(2)}(am_Q)$.

\begin{figure}
\vspace{0.5cm}
\centerline{ \epsfxsize=7.0cm
\epsfbox{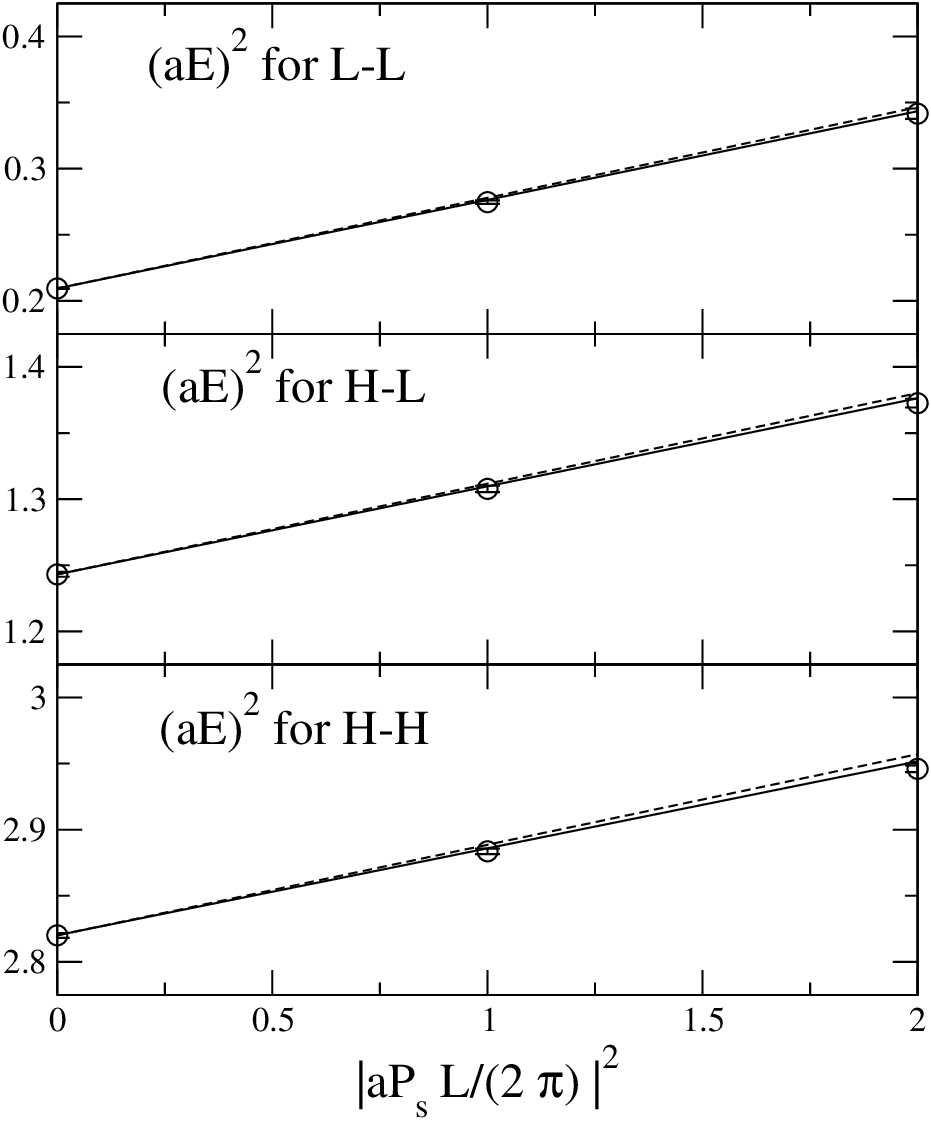}}
\caption{
Momentum dependence of the pseudoscalar meson energies for
the light-light system(top), the heavy-light
system(middle) and the heavy-heavy(bottom) systems 
in case of $\kappa_{heavy}=\kappa_6$ and $\kappa_{light}=\kappa_3$ 
with the Iwasaki gauge action.
The dashed line represents the
continuum dispersion relation with $c_{\rm eff}=1$, while
the solid one represents the fitting results with a linear function.
}
\label{Fig:Dsprsn}
\end{figure}

\begin{figure}
\centerline{ \epsfxsize=14.0cm
\epsfbox{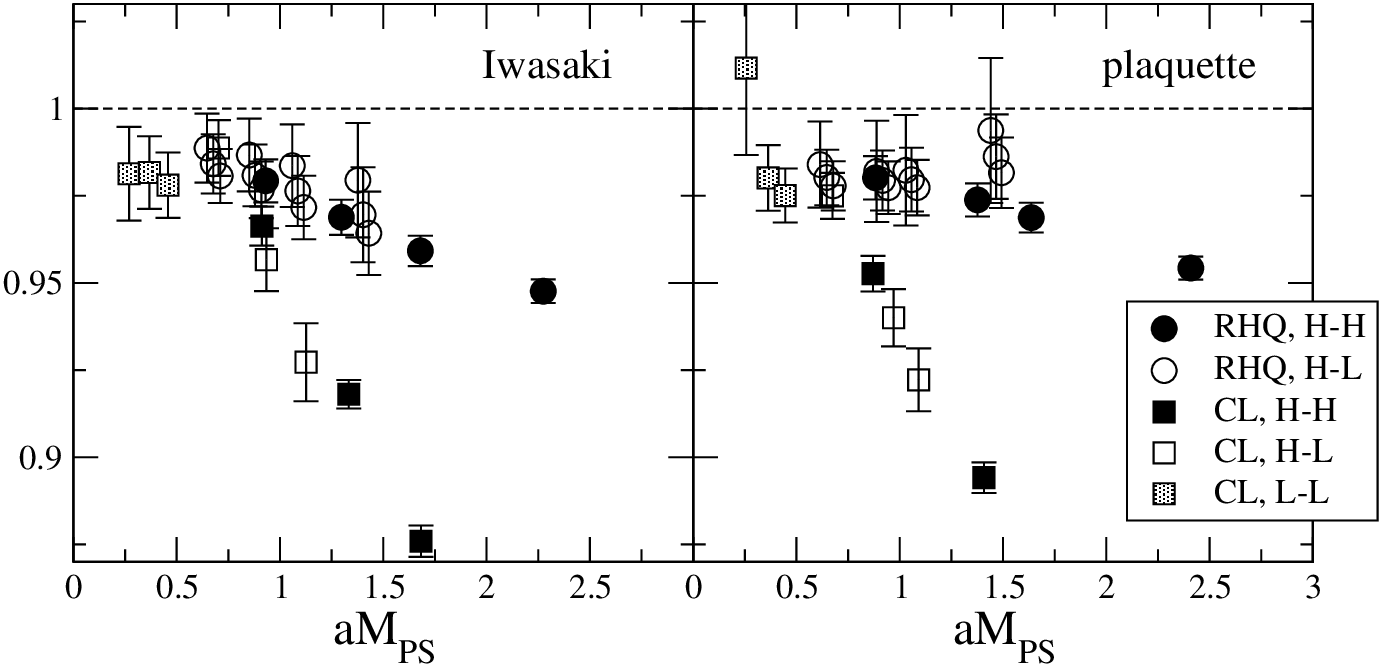}}
\caption{
Effective speed of light for the heavy-heavy and heavy-light pseudoscalar
mesons using the RHQ action and the heavy clover quark action 
with the Iwasaki(left) and the plaquette(right) gauge actions.
}
\label{Fig:SPL.PS}
\end{figure}

\begin{figure}
\vspace{0.5cm}
\centerline{ \epsfxsize=14.0cm
\epsfbox{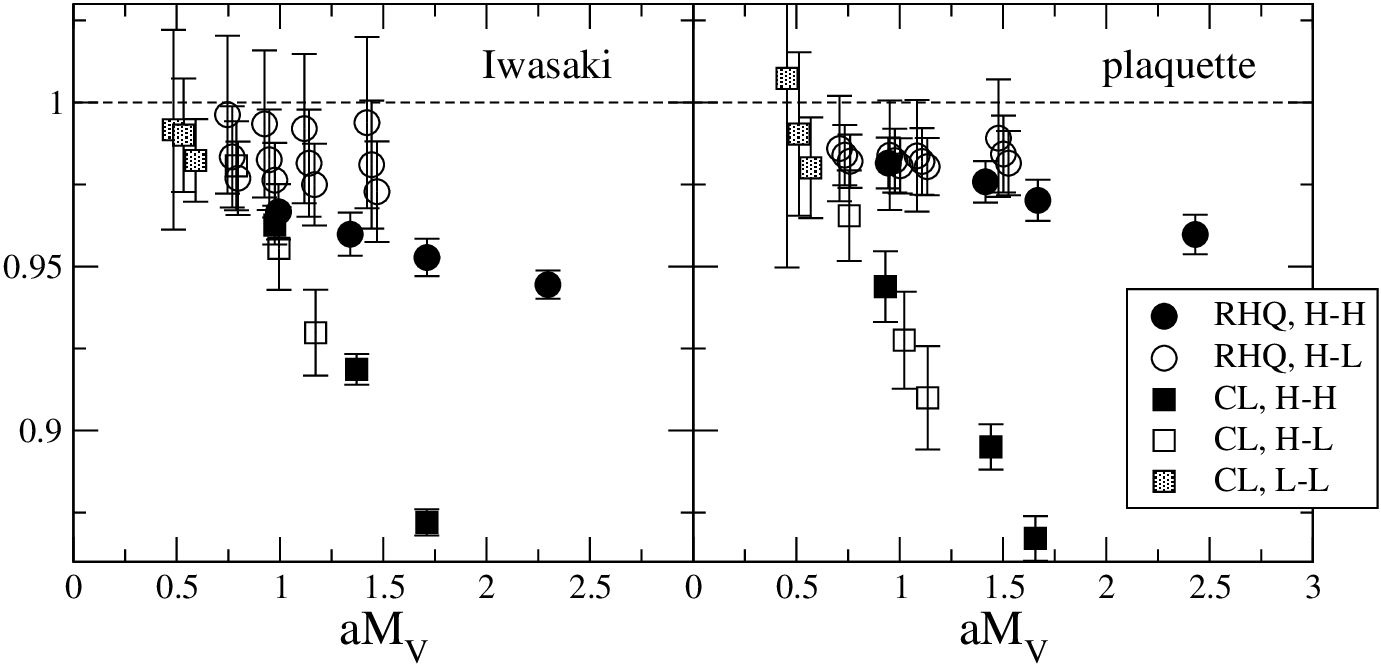}}
\caption{Same as Fig.\ref{Fig:SPL.PS} for the vector
mesons.}
\label{fig:SPL.V}
\end{figure}

We calculate an effective speed of light $c_{\rm eff}$ both for the
pseudo-scalar and vector mesons by
fitting the meson energy $a E({\vec p})$ as a function of the spatial
momentum $a {\vec p}$ with the following form:
\beqa
(a E({\vec p}))^2 = c_{\rm eff}^2 \vert a {\vec p}\vert^2 +(a E({\vec 0}))^2.
\label{Eq:latdsprtn}
\eeqa
In the continuum limit $c_{\rm eff}$ should become unity.
At finite lattice spacing, however, $c_{\rm eff}$ deviates
from unity due to the lattice cutoff errors. 
In Fig.\ref{Fig:Dsprsn} $(a E)^2$ is plotted as a function of 
 $(a{\vec p})^2$, where the fitting result with Eq.~(\ref{Eq:latdsprtn})
is given by the solid line together with the continuum dispersion 
relation with $c_{\rm eff}=1$ represented by the dashed line. 
We observe that the linearity of $E^2$ in $|{\vec p}|^2$ 
is well satisfied and $c_{\rm eff}$ is close to unity.
Fitted values of $c_{\rm eff}$ for the L-L, H-L and H-H cases 
are plotted in Fig.~\ref{Fig:SPL.PS} for the pseudo-scalar mesons and
in Fig.~\ref{fig:SPL.V} for the vector mesons. 
Here it should be noted that in addition to finite quark mass errors
$c_{\rm eff}$ suffers from finite momentum corrections of $O(|a
\vec{p}|^2)$
so that
$c_{\rm eff}$ could deviate from
unity even for the massless quarks. 
Indeed Fig.\ref{Fig:SPL.PS} shows that 
as the meson mass $M a$ decreases,
$c_{\rm eff}$ becomes closer to unity within this uncertainty.
In the heavy quark mass region around $M a \sim 1-2$, $c_{\rm eff}$ for 
the heavy clover quark action deviates from unity by about $7-10\%$.
On the other hand, the RHQ action satisfies $c_{\rm eff}=1$ 
within $2-3\%$ errors,
which are comparable to the deviation for the L-L case.
Since fitted values of $c_{\rm eff}$  for the vector mesons 
in Fig.~\ref{fig:SPL.V} 
are consistent with those for the pseudo-scalar mesons 
within statistical errors, 
we use the values of $c_{\rm eff}$ determined from the pseudo-scalar 
meson dispersion relation in the following discussion.
We observe no obvious difference in the results between the 
Iwasaki and plaquette gauge actions.

\begin{figure}
\vspace{0.5cm}
\centerline{ \epsfxsize=14.0cm
\epsfbox{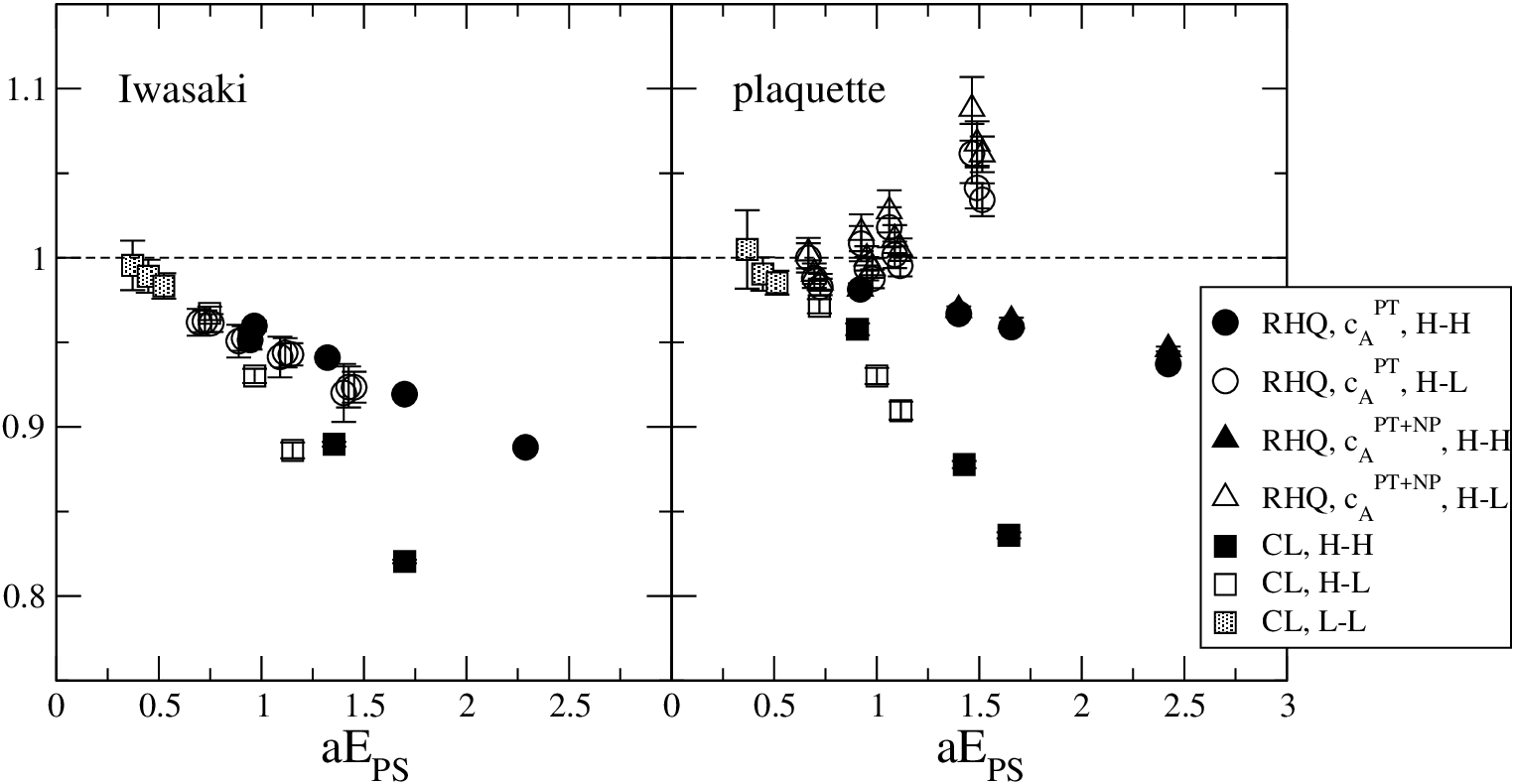}}
\caption{
$R$ function defined by Eq.(\ref{Eq:Rdef}) 
for the heavy-heavy and heavy-light pseudoscalar
mesons using the RHQ action and the heavy clover quark action 
with the Iwasaki(left) and the plaquette(right) gauge actions.
}
\label{fig:AIS}
\end{figure}

\begin{figure}
\centerline{ \epsfxsize=14.0cm
\epsfbox{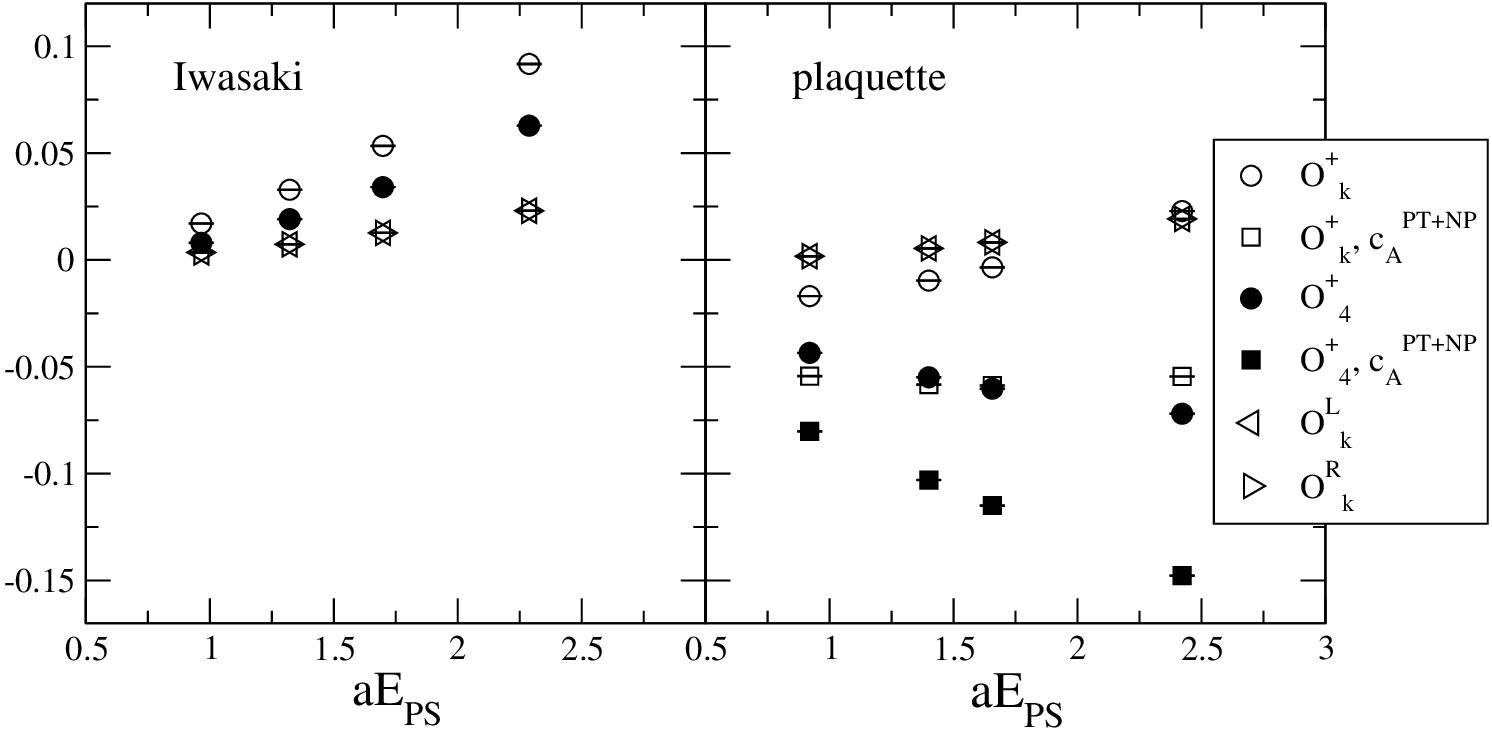}}
\caption{Contribution of each improvement term normalized by the
diagonal one for the heavy-heavy axial vector current.}
\label{fig:AIS.ImprOP.HH}
\end{figure}

\begin{figure}
\vspace{0.5cm}
\centerline{ \epsfxsize=14.0cm
\epsfbox{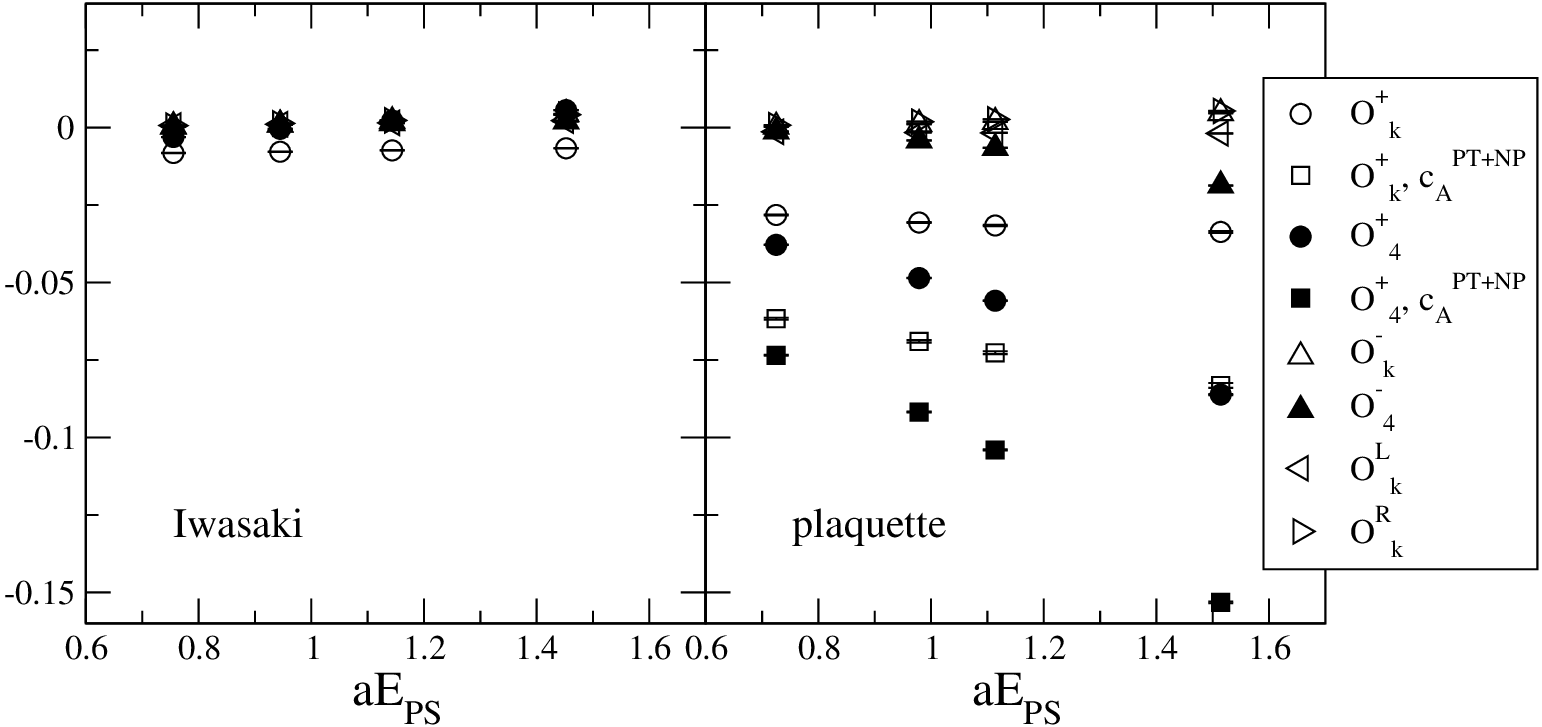}}
\caption{Same as Fig.\ref{fig:AIS.ImprOP.HH} for the
heavy-light axial vector current.}
\label{fig:AIS.ImprOP.HL}
\end{figure}

We also study the space-time symmetry of the pseudo-scalar
meson decay matrix element defined by
\beqa
R \equiv i\frac{\langle 0|A_k^R|PS({\vec p})\rangle}
{\langle 0|A_4^R|PS({\vec p})\rangle}
\frac{E_{PS}}{|p_k|} ,
\label{Eq:Rdef}
\eeqa
where $A_k^R$ and $A_4^R$ represent the spatial and temporal
components of the renormalized axial vector current given in
Eq.(\ref{eq:RHQcrrntImpr}). The pseudo-scalar meson state has
finite spatial momentum of $|a \vec{p}|=2 \pi/24$.
The ratio $R$ is plotted in Fig.\ref{fig:AIS} 
as a function of the meson energy $E_{PS}$ with the lowest finite
spatial momentum for the L-L, H-L and H-H systems, where
$c_A^{\rm PT+NP}$ represents 
the partial replacement of the perturbative value for $c_{A_\mu}^+$
by the nonperturbative  one defined in Eq.(\ref{Eq:cANPshift}),
while  $c_A^{\rm PT}$ means the perturbative value for $c_{A_\mu}^+$ 
without this replacement.
For the plaquette gauge action we employ $Z_{A_\mu}$ 
defined in Eq.(\ref{Eq:zANPshift}), though 
$Z_{A_\mu}^{\rm PT}=0.814$ and $Z_{A_\mu}^{\rm NP}=0.807$ 
agree with each other within $1\%$.
Although because of the finite momentum corrections
the ratio $R$ could deviate from unity even
for the massless quarks, 
it becomes consistent with unity
within the statistical errors as the meson
energy $a E$ vanishes.  
For the massive mesons with $a E \sim 1-2$, on the other hand, 
the heavy clover quark action violates 
the space-time symmetry by about $7-13\%$, 
while the RHQ action retains $R=1$ within $6\%$ errors.
An intriguing observation is that the ratio $R$ of the 
H-L system shows different $aE_{PS}$ dependences  
between the Iwasaki and plaquette gauge actions:
the ratio $R$ decreases for the Iwasaki action
as $E_{PS}$ increases, 
while it increases for the plaquette action.
This different behaviors could come from a fact that the
contributions of the $O(a)$ improvement operators 
are sizable for 
the plaquette action, whereas they are small for the Iwasaki action.
This is observed in
Figs.\ref{fig:AIS.ImprOP.HH} and \ref{fig:AIS.ImprOP.HL}
which show the relative contribution from each $O(a)$ improvement operator
of Eqs.(\ref{eq:A corr term +})-(\ref{eq:A corr term R}) 
to the axial-vector currents defined by
\be
\frac{\sum_{\vec x}{\langle  O_\mu^{\{\pm,L,R\}}({\vec x},t)  P(0)
\rangle }}
{\sum_{\vec x}{\langle A_\mu({\vec x},t) P(0) \rangle}}.
\ee
Dominant contributions always come from $O_\mu^+$ operators 
for the plaquette action,
while their contributions  
are not so large for the Iwasaki action.
In particular, this feature is more prominent for the H-L system.

From the above analyses on $c_{\rm eff}$ and $R$ 
it can be concluded that the RHQ action succeeds in significantly
reducing the $(m_Q a)^n$ errors in the heavy clover quark action.

\subsection{Physical quantities of S-state charmed mesons}
\label{Sec:chrm.spctrm}

\subsubsection{Physical points}
\label{Physical points;Nf=0}

In order to obtain the meson spectra and the decay constants 
at the physical quark masses, we
have to interpolate the heavy quark mass to the charm quark mass $m_{c}$,
while extrapolating the light quark mass
to the $u,d$ quark mass $m_{ud}$ or 
interpolating it to the strange quark mass $m_s$.
Since we employ only 3 values of the light quark masses in our
simulation, we consider only
a linear extrapolation to the $u,d$ quark mass.
In the following the lattice spacing is 
always determined by the Sommer scale with $r_0=0.5$fm.

The light-light pseudo-scalar meson masses are linearly fitted 
in $1/\kappa$ as
\beqa
a^2 M_{PS}^2 =
A\left( \frac{1}{\kappa_{light}}-\frac{1}{\kappa_{crit}}\right),
\label{Eq:Fit:Mpi}
\eeqa
where $\kappa_{crit}$ is determined from the vanishing 
point of $(a M_{PS})^2$. 
$\kappa_{ud}$ and $\kappa_s$ are determined so as to satisfy
$M_{PS}=M_{\pi}=135.0$MeV
and $M_{PS}=M_K=497.7$MeV, respectively.
The fitting results of $A$ and $\kappa_{crit}$ are 
tabulated in Table~\ref{Tab:Fit:Mpi} and
$\kappa_{ud}$ and $\kappa_s$ are given in Table~\ref{tab:PhysP}.

We determine $\kappa_c$ in two different ways:
matching $M_{PS}^{\rm pole}$ to $M_{D_{s}}=1.9683$GeV for the charmed-strange 
meson 
or $M_V^{\rm pole}$ to $M_{J/\psi}=3.0969$GeV for the charmonium, 
where the superscript pole represents a pole mass 
determined from an exponential fall-off of the meson
correlator. 
Employing the following fitting functions
\beqa
a M_{PS}^{\rm pole} =
A+B\kappa_{heavy}+C\kappa_{heavy}^2+D
a m_q^{light}
\label{Eq:Fit:Mhl}
\eeqa
for the heavy-light meson masses with 
$ a m_q^{light}=(1/\kappa_{light} -1/\kappa_{crit})/2$
and
\beqa
a M_V^{\rm pole} = A+B\kappa_{heavy}+C\kappa_{heavy}^2
\label{Eq:Fit:Mhh}
\eeqa
for the heavy-heavy meson masses, 
we have determined two values of $\kappa_{c}$,   
which are given in Table~\ref{tab:PhysP}.

In order to estimate a magnitude of the cutoff errors,
we also calculate the charmed meson spectra
employing the kinetic mass defined by 
\beqa
a M^{\rm kin} =a M^{\rm pole}/c_{\rm eff}^2.
\label{Eq:Mkin:def}
\eeqa
With the same fitting functions as Eqs.(\ref{Eq:Fit:Mhl}) and
(\ref{Eq:Fit:Mhh}) we have also determined 
$\kappa_c(D_s, M^{\rm kin})$
and
$\kappa_c(J/\psi, M^{\rm kin})$ listed in Table~\ref{tab:PhysP}.
From these results we observe that
a difference of $\kappa_c$ between two
physical inputs $M_{J/\psi}$ and $M_{D_s}$ 
is less than $0.2\%$, while
a difference of $\kappa_c$  between $M^{\rm pole}$ 
or $M^{\rm kin}$ is about $2\%$. 
In the following analysis, 
we always calculate all the physical quantities using both $M^{\rm
pole}$ and $M^{\rm kin}$,
in order to estimate the systematic
errors due to an ambiguity in the choice
of $M^{\rm pole}$ or $M^{\rm kin}$.

\subsubsection{Hyperfine splitting for charmonium and charmed-strange meson}

\begin{figure}
\centerline{ \epsfxsize=14.0cm
\epsfbox{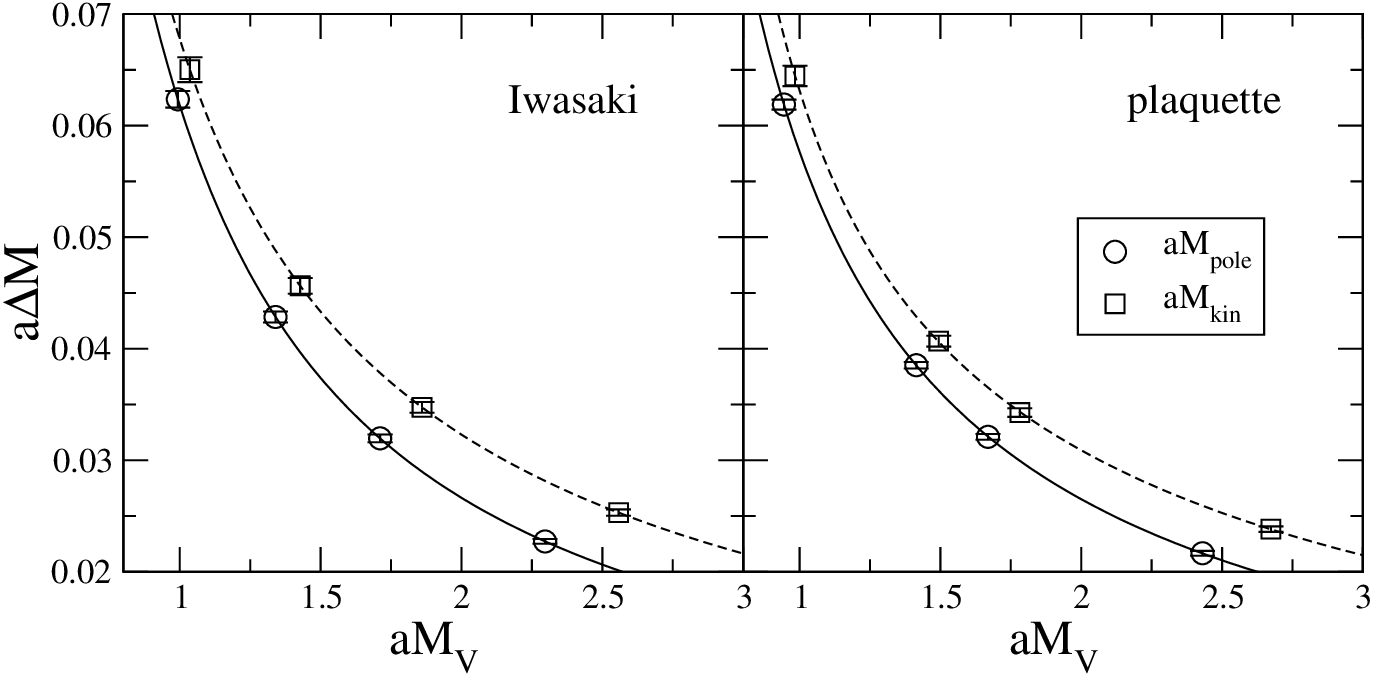}}
\caption{Heavy-heavy meson S-state hyperfine splittings as a
function of $M_V a$.}
\label{fig:Fit.CC.HFS}
\end{figure}

Figure~\ref{fig:Fit.CC.HFS} shows $aM_V^X $ dependence of
the S-state charmonium hyperfine splitting 
$a\Delta M^X = aM_V^X  - aM_{PS}^X $, where $X=$ pole or kin.
In order to interpolate the results at the physical charm quark mass,
we adopt the ansatz that the splitting is 
a polynomial of the inverse vector meson mass:
\beqa
a\Delta M^X = A/(aM_V^X ) +B/(aM_V^X )^2 +C/(aM_V^X )^3,
\label{Eq:Fit:CCHFS}
\eeqa
incorporating a property that the hyperfine splitting vanishes 
in the infinite quark mass limit
due to the heavy quark symmetry.
The interpolation lines are also plotted in Fig.\ref{fig:Fit.CC.HFS}. 
Using the fitting results for the parameters $A$, $B$ and $C$ 
given in Table~\ref{Tab:Fit:Mcc},
we obtain $\Delta M(J/\psi-\eta_c)$ in physical unit.
$\Delta M^{\rm pole}$ at $\kappa_c(J/\psi, M^{\rm pole})$ and
$\Delta M^{\rm kin}$ at $\kappa_c(J/\psi, M^{\rm kin})$ are tabulated in
Table~\ref{Tab:HFS:CC} for each gauge action 
together with the experimental value.

\begin{figure}
\vspace{0.5cm}
\centerline{ \epsfxsize=14.0cm
\epsfbox{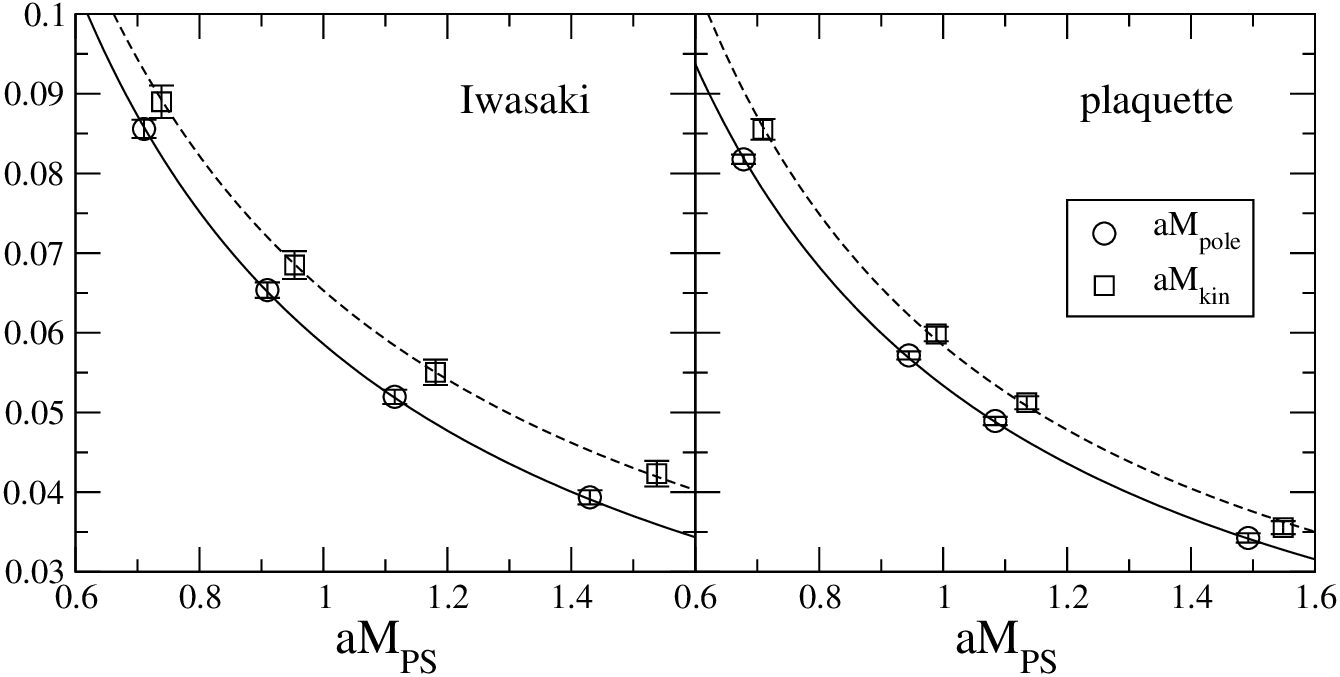}}
\caption{
Heavy-light meson S-state hyperfine splittings as a function of
$aM_{PS}$ with $\kappa_{light}=\kappa_3$.
}
\label{fig:Fit.CS.HFS}
\end{figure}

In Fig.\ref{fig:Fit.CS.HFS} we plot
the S-state charmed-strange meson hyperfine splitting
$a\Delta M^X = aM_V^X  -a M_{PS}^X $ as a function
of $aM_{PS}^X $ together with the interpolation lines
which are obtained by
employing the ansatz motivated by the heavy quark symmetry:
\beqa
a \Delta M^X =(A +B a m_q^{light})/(a M_{PS}^X) +C/(a
M_{PS}^X)^2 .
\label{Eq:FitCSHFS}
\eeqa
Using the fitting results presented in Table~\ref{Tab:Fit:HFS:CS},
we obtain $\Delta M(D_s^*- D_s)$ in physical unit.
$\Delta M^{\rm pole}$ at $\kappa_c(D_s, M^{\rm pole})$ and $\kappa_s$,
and
$\Delta M^{\rm kin}$ at $\kappa_c(D_s, M^{\rm kin})$ and $\kappa_s$
are listed  in Table~\ref{Tab:HFS:CC} for each gauge action 
together with the experimental value.

\subsubsection{$D_s$ meson decay constants}

\begin{figure}
\centerline{ \epsfxsize=14.0cm
\epsfbox{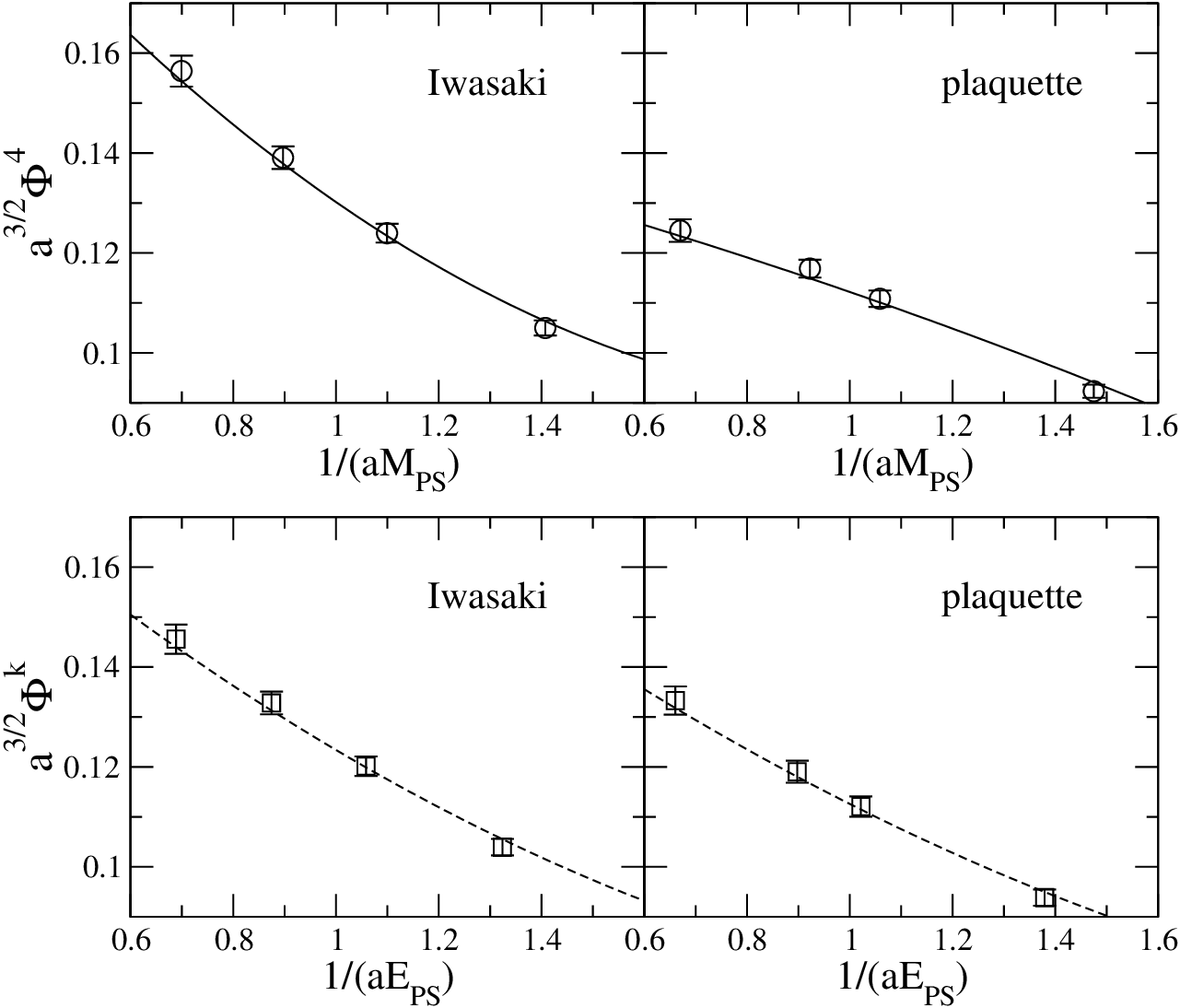}}
\caption{
Charmed-strange pseudoscalar meson decay constants obtained from
the temporal(top) and spatial(bottom) 
components of the axial vector current
as a function of $a M_{PS}$ in case of $\kappa_{light}=\kappa_3$.}
\label{fig:Fit.CS.fDs}
\end{figure}

The heavy-light pseudo-scalar meson decay constant $a f_{PS}$ can be
obtained from the temporal and spatial components 
of Eq.(\ref{Eq:fPS:def}).
In our calculation $f_{PS}(A_4)$ is determined from 
\beqa
a^{3/2} \Phi_P^{4} \equiv \sqrt{a M_{PS}^{\rm pole}}~a f_{PS}
= {Z_{A^{\rm impr}_{\bf 4}}}/\sqrt{a M_{PS}^{\rm pole}}
\label{Eq:fPS:drvtn4}
\eeqa
and $f_{PS}(A_k)$ from 
\beqa
a^{3/2} \Phi_P^{k} \equiv \sqrt{a E_{PS}^{\rm pole}}~a f_{PS}
= {Z_{A^{\rm impr}_{\bf k}}} \sqrt{a E_{PS}^{\rm pole}}/(i a
p_k),
\label{Eq:fPS:drvtnk}
\eeqa
where ${Z_{A^{\rm impr}_{\bf 4}}}$ and 
${Z_{A^{\rm impr}_{\bf k}}}$ are the decay matrix
elements defined in Eq.(\ref{eq:zaimpr matrix el}).
Note that only the improved axial vector current
with $c_{A_\mu}^{\rm PT+NP}$ is considered
for the plaquette gauge action. In Fig.\ref{fig:Fit.CS.fDs}
we plot
$a^{3/2} \Phi_P^4$ and $a^{3/2} \Phi_P^k$ as a function of
$1/(aM_{PS}^{\rm pole, kin})$. 
The interpolation lines are obtained 
by fitting the results with the following ansatz:  
\beqa
a^{3/2} \Phi_{PS}^4 &=& A +B/(aM_{PS}^{\rm pole, kin}) 
+C/(aM_{PS}^{\rm pole,kin})^2
+D a m_q^{light}  ,\\
a^{3/2} \Phi_{PS}^k &=& A +B/(aE_{PS}^{\rm pole, kin}) +C/(aE_{PS}^{\rm pole,kin})^2
+D a m_q^{light}  .
\label{Eq:Fit:fCS}
\eeqa
Using the fitted values of the parameters 
in Table~\ref{Tab:Fit:fcsA4}-\ref{Tab:Fit:fcsAk},
we obtain $f_{D_s}$ in physical unit. Table~\ref{Tab:fCS} lists
the results of $f_{PS}$ at $\kappa_c(D_s, M^{\rm pole})$ and $\kappa_s$
and $f_{PS}$ at $\kappa_c(D_s, M^{\rm kin})$ and $\kappa_s$
for each gauge action together with the experimental value.


\subsection{Cutoff effects}
\label{Sec:Systematic errors}

We now consider the cutoff effects in our results.
Leading cutoff effects for the gauge part
are $O(a^2 \lqcd^2)$. 
The light quark action also has $O(a^2 \lqcd^2)$ errors,
since the nonperturbative value of $c_{\rm SW}$ is employed
for each gauge action.
For the RHQ action, on the other hand, the leading cutoff effects are
$O(\alpha_s(\mu)^2 f_0^{(2)}(a m_Q))$ with $\alpha_s(\mu)=g^2(\mu)/(4 \pi)$, 
which comes from 
the fact that the parameter $\nu$ associated with the $O(1)$ 
kinetic term is only adjusted up to one-loop level. 
Since this error is responsible for the deviation of $c_{\rm eff}$  from unity,
the mass dependence of $c_{\rm eff}$ shown 
in Figs.\ref{Fig:SPL.PS} and \ref{fig:SPL.V}
tells us that $f_0^{(2)}(m_Q a)$ is a smooth function of $a m_Q$ 
in the range of the heavy quark mass employed in our simulation.
In addition, 
there exists the $O(\alpha_s(\mu)^2 g_0^{(2)}(a m_Q))$ errors originating from
the heavy quark axial vector currents whose renormalization factors are
determined up to one-loop level.
These are the leading cutoff effects in
the deviation of $R$ from unity shown in Fig.\ref{fig:AIS},
where we find fairly smooth $a m_Q$ dependence.
  
Let us take into account 
these $O(\alpha_s(\mu)^2 f_0^{(2)}(a m_Q),
\alpha_s(\mu)^2 g_0^{(2)}(a m_Q))$  effects in our error estimate
using a difference of the charmed meson hyperfine splittings
obtained with $M_{\rm pole}$ and $M_{\rm kin}$ and
also a difference of the charmed meson decay constants
extracted from $A_k$ and $A_4$. 
For the hyperfine splittings we take the pole mass result 
as the central value and
a difference between two results as a systematic error.
In Table~\ref{Tab:HFS:CC:Fnl} 
our final result for the charmonium hyperfine splitting
in physical unit is also presented, where the central value is 
$\Delta M_{\rm pole}(J/\psi-\eta_c)$, 
the first error is statistical and the second is 
a systematic error explained above.
The second error, much larger than the first, is about $16\%$ for the Iwasaki
action and about $12\%$ for the plaquette action.
Similarly, our final result for the charmed-strange meson hyperfine
splitting in physical unit is given in Table\ref{Tab:HFS:CC:Fnl}, 
where the central value is
$\Delta M_{\rm pole}(D_s^*-D_s)$, 
the first error is statistical and 
the second is a systematic error.
It is interesting that 
the second errors for the charmed-strange meson hyperfine
splitting , which are about $8\%$  for the Iwasaki
action and about $7\%$ for the plaquette action, 
are half of those for the charmonium hyperfine splitting. 
This suggests that the dominant systematic errors come from
the heavy quarks, so that they are proportional to 
a number of heavy quarks in the mesons.
In Table~\ref{Tab:HFS:CC:Fnl}
our final result for the $D_s$ meson decay constant 
in physical unit is also presented, where we take $f_{D_s}(A_4)$ 
with $M_{\rm pole}$ as the central value.
The first error is statistical and the second and the third are 
systematic errors
estimated from a difference of $f_{D_s}(A_4)$ 
between $M_{\rm pole}$ and $M_{\rm kin}$
and  a difference between $f_{D_s}(A_4)$
and $f_{D_s}(A_k)$ with $M_{\rm pole}$, respectively.
Both the second and third errors are less than $1\%$ 
for the plaquette gauge action.
For the Iwasaki gauge action, on the other hand,
the third error is about $5\%$ though
less than $1\%$ for the second.
Smallness of the third error for the plaquette action 
may be partly due to the use of $c_A^{\rm PT+NP}$.
Note that the systematic errors associated with the heavy quark action are
estimated at one lattice spacing in this paper. Therefore, in future
works, it is desirable to study these systematic errors by
changing the lattice spacing.

Once the systematic errors are taken into account,
our results of the hyperfine splitting for two gauge actions 
agree with each other.
For $f_{D_s}$, on the other hand, an agreement is not so excellent:
the difference is still 1.5$\sigma$ even if we take the 
systematic error for the Iwasaki action. 
It could be interesting to see whether the difference diminishes
if we employ $c_A^{\rm PT+NP}$ for the Iwasaki gauge action.


\subsection{Comparison with the previous results}
\label{Sec:Comparison}

\begin{figure}
\centerline{ \epsfxsize=12.0cm
\epsfbox{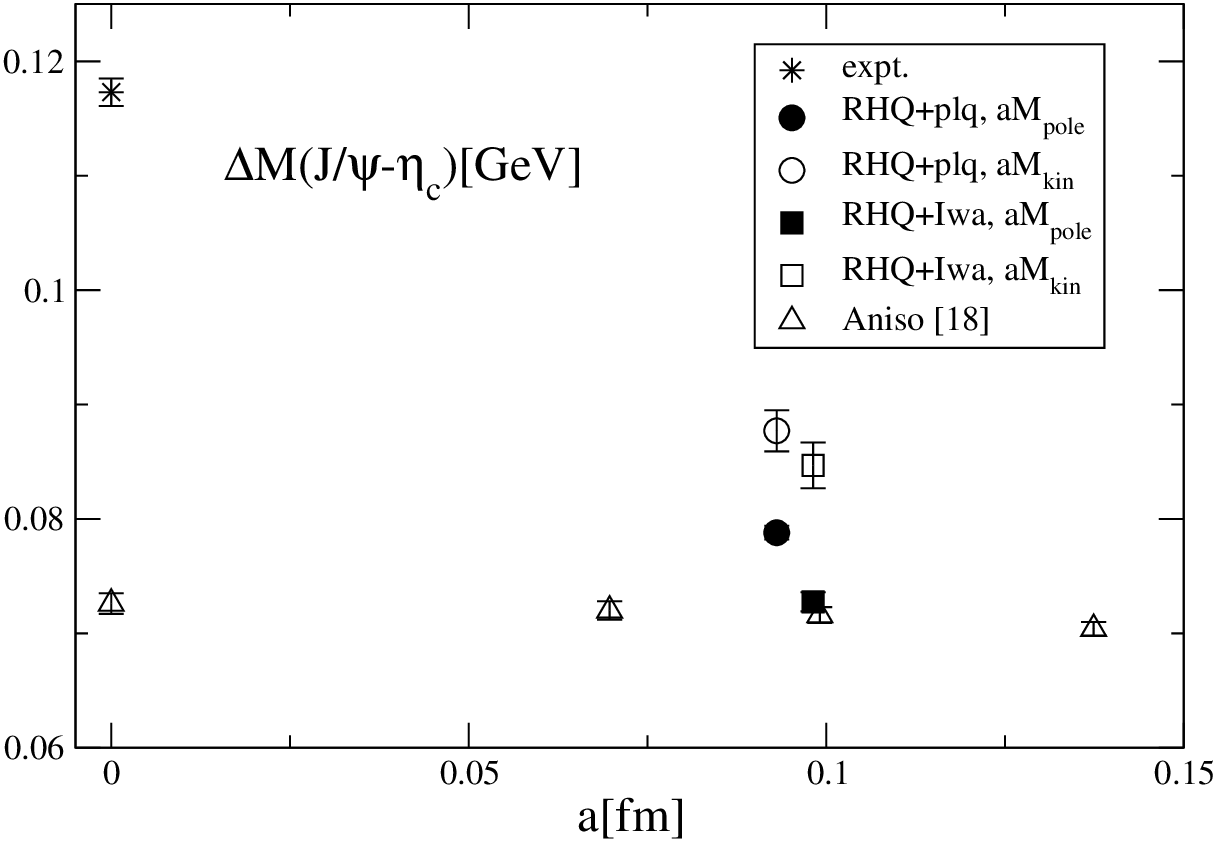}}
\caption{
Comparison of charmonium S-state hyperfine splitting in physical unit.}
\label{fig:HFS.r0}
\end{figure}

In Fig.\ref{fig:HFS.r0} our results of the S-state
charmonium hyperfine splitting are compared 
with a previous result obtained 
by the CP-PACS collaboration
using the anisotropic lattice QCD\cite{Aniso:CC:Spctrm}, where
the effective speed of light is nonperturbatively 
adjusted to unity such that 
$M^{\rm pole} = M^{\rm kin}$. Both results are plotted as a function of 
the lattice spacing determined by the Sommer scale $r_0=0.5$fm.
Our result with the pole mass for the Iwasaki gauge action is consistent with
the continuum limit of the anisotropic lattice result 
within the small statistical error,
though the kinetic mass result is rather large.
For the plaquette gauge action,
on the other hand, both the pole and kinetic mass results are larger than  
the anisotropic lattice results.
The large systematic error due to the pole to kinetic mass difference 
should be eliminated with the 
use of nonperturbative $\nu$ in future calculations.
It should be noted that all the results are
smaller than the experimental value  by  about $40\%$.

\begin{figure}
\vspace{0.5cm}
\centerline{ \epsfxsize=12.0cm
\epsfbox{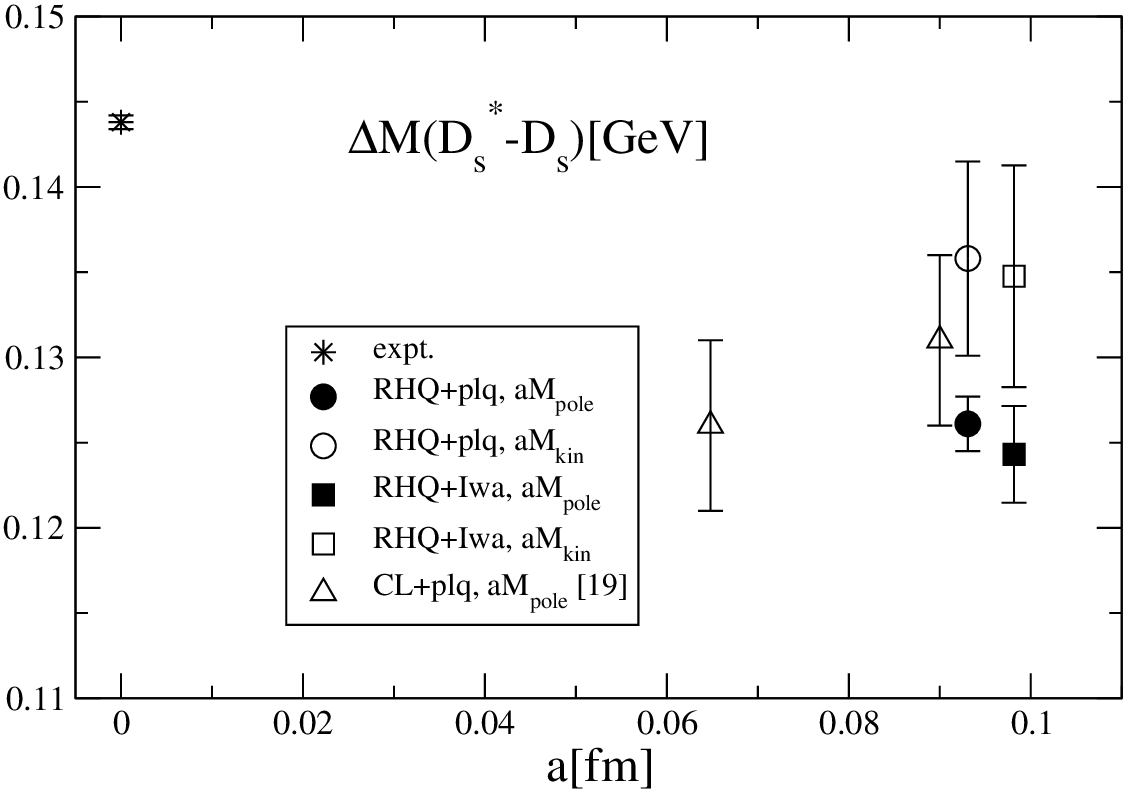}}
\caption{
Comparison of $D_s$ meson hyperfine splitting in physical unit.
The results of Ref.\cite{fDs.UKQCD} are slightly shifted in $a$
coordinate for visibility.}
\label{fig:SP.r0}
\end{figure}

Figure~\ref{fig:SP.r0} shows the comparison of 
our results of the S-state charmed-strange meson
hyperfine splitting with a previous result
obtained by the UKQCD collaboration 
using the heavy clover quark action\cite{fDs.UKQCD}.
We observe that all the results agree within large
statistical errors, though they are smaller than the experimental value 
by about $10\%$.

\begin{figure}
\centerline{ \epsfxsize=12.0cm
\epsfbox{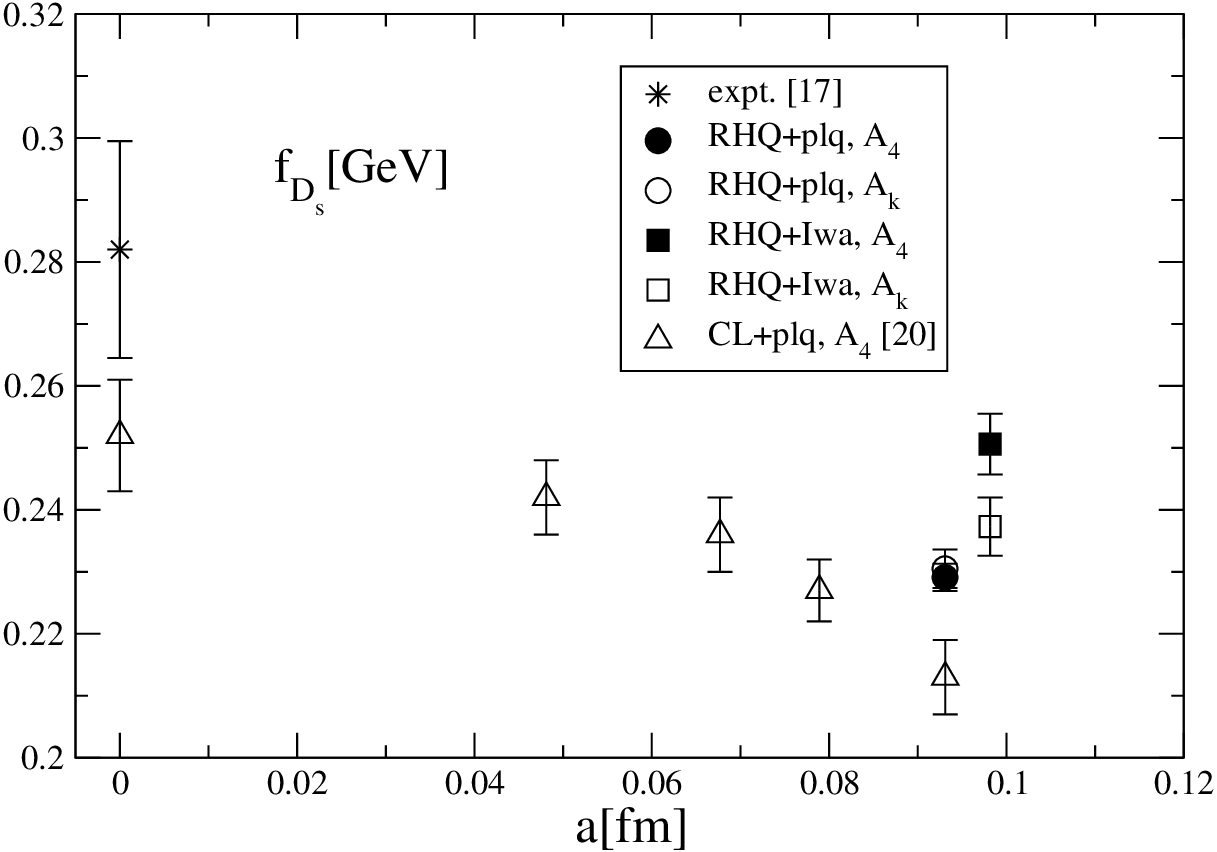}}
\caption{Comparison of $D_s$ meson decay constant in physical unit.}
\label{fig:fDs.r0}
\end{figure}

In Fig.\ref{fig:fDs.r0} we compare our results of $f_{D_s}$ with
a previous result obtained by 
the ALPHA collaboration
using the heavy clover quark and the plaquette gauge actions\cite{fDs.ALPHA}.
Our results at finite lattice spacing are closer to the
ALPHA result at the continuum limit than at a similar lattice spacing.
This could indicate that $f_{D_s}$ from the RHQ action 
has a good scaling behavior,
which should be checked in future scaling studies.
We also point out that $c_A^{\rm PT+NP}$ for the Iwasaki gauge action may 
reduce the difference between $f_{D_s}(A_4)$ and $f_{D_s}(A_k)$.
We also leave it to future work.


\section{Conclusion}
\label{Sec:Conclusions}

We have carried out a first nonperturbative test of the RHQ action 
focusing on the magnitude of the cutoff errors.
We investigate 
the dispersion relation of the pseudo-scalar meson 
and the space-time symmetry for the
pseudo-scalar meson decay matrix element.
Our results show that the RHQ action has much smaller
cutoff errors than the heavy clover quark action
around the charm quark mass.

We also investigate the systematic errors due to the cutoff effects 
for the physical observables.
In case of the charmonium (charmed-strange) hyperfine splitting,
a difference between the results with $M^{\rm pole}$ and $M^{\rm kin}$ 
is used to estimate
the systematic error, which is as large as $16\%$ ($8\%$) 
for the Iwasaki gauge action and $12\%$ 
($7\%$) for the plaquette gauge action. 
For the $D_s$ meson decay constant $f_{D_s}$,
we estimate the systematic error by a difference between $f_{D_s}(A_4)$ and
$f_{D_s}(A_k)$ as well as 
a difference between $M^{\rm pole}$ and $M^{\rm kin}$.
The latter is negligible for both gauge actions, while 
the former is about $5\%$ for the Iwasaki gauge
action and $0.5\%$ for the plaquette gauge action.

There are two important subjects for future studies.
One is a further improvement of the RHQ action to reduce the cutoff effects.
In particular, it is rather easy to tune the improvement 
coefficient $\nu$ nonperturbatively, which is supposed to
eliminate the leading $O(\alpha_s^2)$ errors. 
This study is under way\cite{latt05}.
The other is the inclusion of light dynamical quark effects.
It is interesting to investigate whether the deficit in  
the quenched value for the S-state charmonium hyperfine splitting 
is fully accounted by the sea quark effects.


\section*{Acknowledgments}
This work is supported in part by Grants-in-Aid for Scientific Research 
from the Ministry of Education, Culture, Sports, 
Science and Technology (Nos.~13135204,
13640259, 
13640260, 
15204015, 
15540251, 
15540279, 
15740165, 
16028201, 
16540228, 
17340066 
18540250 
).

%
\appendix

\section{Renormalization factors and improvement
coefficients for massive quarks}
\label{appndx:A}

In this appendix we explain how to determine the input parameters 
for the RHQ action and the axial vector currents in our numerical simulation,
such as $\kappa$,  the improvement coefficients and the renormalization factors
together with the mean field improvement discussed in Refs.\cite{cEcB,zAcA}.

The mean field improvement is introduced as the 
redefinition of link variable $U_\mu(x) \rightarrow u_0
(U_\mu(x)/u_0) \equiv u_0 \tilde{U}_\mu(x)$,
where  $u_0=P^{1/4}$ with the averaged plaquette value $P$ in our simulation.
The one-loop expression for $u_0$ is given by
\be
u_0 = 1 -g^2 \frac{C_F}{2} T_{\rm MF}, 
\ee
where $T_{\rm MF}=1/8$ for the plaquette gauge action and
$0.0525664$
for the Iwasaki gauge action\cite{Tadpole.val}.

With the replacement $U_\mu(x) \rightarrow u_0
\tilde{U}_\mu(x)$ it is natural to introduce the boosted gauge
coupling $g_0^2/u_0^4$, which is related to the
$\msbar$ coupling constant $g_{\msbar}^2 (\mu)$
with the scale $\mu=1/a$ as
\be
\frac{1}{g^2_\msbar(\mu)} = 
\frac{c_0 P+8 c_1 R}{g_0^2}-0.1006 +0.03149 N_f +
\frac{11-\frac{2}{3}N_f}{8\pi^2} {\rm log}(a \mu)
\ee
for the Iwasaki gauge action and the $O(a)$ improved Wilson
quark action\cite{CP-PACS:fHL:NRQCD:Nf=2}, and 
\be
\frac{1}{g^2_\msbar(\mu)} =
\frac{P}{g_0^2}-0.1349 +0.03149 N_f +
\frac{11-\frac{2}{3}N_f}{8\pi^2} {\rm log}(a \mu)
\ee
for the plaquette gauge action and the $O(a)$ improved
Wilson quark action\cite{gmass}.
In the following we simply use $g^2$ to express
$g_{\msbar}^2 (\mu=1/a)$.

The inverse quark propagator at the leading order without the mean-field
improvement is given by
\beqa
a S_q^{-1} &=& a m_0 + g^2 a m_c^{(1)}+i\gamma_4 \sin (p_4 a) + i\nu
\sum_k \gamma_k \sin (p_k a) + r_t(1-\cos(p_4a)) + r_s\sum_k(1-\cos (p_k a))
\eeqa
where $a m_0$ is the bare quark mass appeared in the action. Note that
we include the one-loop contribution to the critical quark mass, $g^2
m_c^{(1)}$, in the leading order. A reason for this will become clear
later. The pole mass $a \mplo$,  determined from the zero of the inverse
propagator by setting $ p_4 = i \mplo$ and $p_k =0$, satisfies
\beqa
\sinh (a \mplo) + r_t \cosh (a \mplo) = a m + r_t ,
\label{eq:qprop dsptn rel0}
\eeqa
where $a m$ is a shifted quark mass defined by
$a m = am_0 + a g^2 m_c^{(1)}$.
If we perform the replacement $U_\mu(x) \rightarrow u_0
\tilde{U}_\mu(x)$ in the RHQ action given in
Eq.(\ref{eq:action_q}) or Eq.(\ref{eq:RHQ action}),
the inverse quark propagator at the leading order with the mean-field
improvement becomes
\beqa
a \tilde S_q^{-1} &=& a m_0 + g^2 a \Delta m_c^{(1)}+i\gamma_4 u_0 
\sin (p_4 a) + i\nu \sum_k \gamma_k u_0 \sin (p_k a) 
+ r_t(1-u_0 \cos(p_4a)) + r_s\sum_k (1-u_0 \cos (p_k a)) \nn \\
\eeqa
where $g^2 a\Delta m_c^{(1)} =  g^2 a m_c^{(1)}-(r_t+3 r_s)(1-u_0)$.
Then the pole mass $a \mplomf$ at the tree-level 
with the mean-field improvement satisfies
\beqa
u_0 \sinh (a \mplomf) 
&=& a m_0 + ag^2 \Delta m_c^{(1)} + r_t (1-u_0 \cosh (a \mplomf))+ 3 r_s(a
\mplomf)(1-u_0) \nn\\
&=& a m + r_t u_0 (1-\cosh(a \mplomf))+(1-u_0)3(r_s(a
\mplomf)-1). 
\label{eq:qprop dsptn rel}
\eeqa
Note that the shifted quark mass $a m $ is kept equal with and without
the mean field improvement. Therefore both $\mplo$ and $\mplomf$ vanish
at $am =0$. Since the remaining one-loop correction to the quark mass is
multiplicative to $m$, the pole masses in both definitions vanish at
$ a m =0$ also at one-loop level.
The inclusion of $ g^2 a m_c^{(1)}$ or $ g^2 a\Delta m_c^{(1)}$ at
leading order is necessary to satisfy this property.
Although in this work we follow the mean-field improvement procedure 
given in Sec.6 of Ref.\cite{cEcB} which does not 
include the $\Delta m_c$ correction, 
the effects on the improvement parameters are less than 1\%.
Eqs.(\ref{eq:qprop dsptn rel0}) and (\ref{eq:qprop dsptn rel}) lead to
the following relation that
\beqa
a\mplo 
&=& a\mplomf +
(u_0-1)\frac{\sinh(a\mplomf)+r_t(\cosh(a\mplomf)-1)
+3 (r_s (a\mplomf)-1) }{\cosh (a\mplomf)+
r_t\sinh(a\mplomf)} \nn\\
&\equiv& {a\mplomf} + g^2 a \Delta m_p,
\eeqa
where
\beqa
a\Delta m_p 
&=& -\frac{ C_F}{2} T_{\rm MF}
\frac{\sinh(a\mplomf)+r_t(\cosh(a\mplomf)-1)
+3 (r_s(a\mplomf)-1)} {\cosh (a\mplomf)+
r_t\sinh (a\mplomf)}.
\eeqa
As a consequence, the quark pole mass is written at the
one-loop level as
\be
a m_p = a\mplo + g^2 a m_p^\nlo = {a \mplomf}+g^2 a
\tilde m_p^\nlo, 
\ee
where $a \tilde m_p^\nlo =a m_p^\nlo +a \Delta m_p$, and $a m_p^\nlo $
is the one-loop correction to the pole mass without the mean field
improvement\cite{cEcB}.

The mean-field improved parameters $Z_q$, $\nu$,
$r_s$, $c_E$ and $c_B$ are given below with the use of
$a \mplomf$ and $a \tilde m_p^\nlo$:
\beqa
Z_{q,\latt}(a\mplomf) 
&=& Z_{q,\latt}^\lo u_0 
\left(1+g^2\frac{Z_{q,\latt}^\nlo}{Z_{q,\latt}^\lo}
+g^2 \frac{C_F}{2}T_{\rm MF}
+\frac{g^2}{Z_{q,\latt}^\lo}
\frac{\p Z_{q,\latt}^\lo}{\p\mplo} {a \Delta m_p}
\right), \\
\nu(a\mplomf) &=&\nu^\lo+g^2\nu^\nlo
+g^2\frac{\partial \nu^\lo}{\partial
a\mplo}{a \Delta m_p},\\
r_s(a\mplomf) &=& r_s^\lo+g^2
r_s^\nlo
+g^2\frac{\partial r_s^\lo}{\partial a
\mplo}{a \Delta m_p},\\
\ce(a\mplomf) &=&\ce^\lo \frac{1}{u_0^3}
\left(1+g^2\frac{\ce^\nlo}{\ce^\lo}
-g^2\frac{3}{2}C_FT_\mf 
+\frac{g^2}{\ce^\lo}
\frac{\p \ce^\lo}{\p \mplo}{a \Delta m_p} \right),\\
\cb(a\mplomf) &=&\cb^\lo
\frac{1}{u_0^3}\left(1+g^2\frac{\cb^\nlo}{\cb^\lo}
-g^2\frac{3}{2}C_FT_\mf
+\frac{g^2}{\cb^\lo}\frac{\p \cb^\lo}
{\p \mplo}{a \Delta m_p} \right),
\eeqa
where one-loop corrections, $Z_{q,\latt}^\nlo$, $\nu^\nlo$, $r_s^\nlo$,
$\ce^\nlo$ and $\cb^\nlo$, have already been calculated in ref.~\cite{cEcB}.
We replace a perturbative value of $u_0$ in the above expressions
by $u_0=P^{1/4}$ with $P$ taken from our simulation. 
We finally determine $\kappa$ in terms of $\mplomf$ as follows.
Using the relation of Eq.(\ref{eq:qprop dsptn rel}) with $r_t=1$
\be
a m_0 +a\Delta m_c = u_0 e^{a \mplomf} -1 -3 r_s (1-u_0),
\ee
the hopping parameter $\kappa$ is given in terms of $\mplomf$:
\be
\kappa \equiv \frac{1}{2}\frac{1}{1+3 r_s+a m_0}
=\frac{1}{2}\frac{1}{u_0 (e^{a \mplomf}+3 r_s)-a\Delta m_c}.
\ee
With this definition, $\kappa$ becomes $\kappa_{crit}$ at the one-loop level
for $a \mplomf=0$.

In a similar manner we can derive 
the renormalization factor and the $O(a)$ improvement
coefficients for the axial-vector currents in Eq.(\ref{eq:RHQcrrntImpr}).
The matching factor $Z_{A_\mu}$ from the lattice to
the continuum $\msbar$ scheme is given in Ref.\cite{zAcA}:
\beqa
Z_{A_\mu}=
\frac{Z_{A_\mu}^\latt}{Z_{A_\mu}^{\msbar}}
&=&\sqrt{Z_{Q,\latt}^\lo(a \mphlomf)}
\sqrt{Z_{q,\latt}^\lo(a
\mpllomf)}u_0\left(1-g^2\Delta_{A_\mu}\right.
\nn\\
&&\left.
+g^2\frac{C_F}{2}T_{\rm MF}
+\frac{1}{2}\frac{g^2}{Z_{Q,\latt}^\lo}
\frac{\partial Z_{Q,\latt}^\lo}{\partial \mphlo}{a
\Delta \mph}
+\frac{1}{2}\frac{g^2}{Z_{q,\latt}^\lo}
\frac{\partial Z_{q,\latt}^\lo}{\partial \mpllo}{a
\Delta \mpl}
\right),
\eeqa
where $\Delta_{A_\mu}$ is the one-loop correction to the renormalization
factor of $A_\mu$\cite{zAcA}.
For the $O(a)$ improvement coefficients, on the other hand,
we use the expressions of Eq.(\ref{eq:RHQcrrntImpr}) suitable
for our numerical simulations, which are related to those
in Ref.\cite{zAcA} as
\beqa
c_{A_k}^+ &=& g^2 c_{A_k}^{+,{\rm PT}}, \\
c_{A_4}^+ &=& g^2 c_{A_4}^{+,{\rm PT}}
(a \mqlomf +a \mQlomf)/
{\rm sinh}(a \mqlomf+a \mQlomf),
\label{eq:redefine cA4+}\\
c_{A_\mu}^- &=& g^2 c_{A_\mu}^{-,{\rm PT}}/u_0,
\label{eq:redefine cA-}\\
c_{A_\mu}^L &=& - g^2 c_{A_\mu}^{L,{\rm PT}}/u_0,
\label{eq:redefine cAL}\\
c_{A_\mu}^R &=& g^2 c_{A_\mu}^{R,{\rm PT}}/u_0 ,
\label{eq:redefine cAR}
\eeqa
where $c_{A_\mu}^{\{\pm,L,R\},{\rm PT}}$ are calculated as
a function of $\mQlomf$ and the superscript
PT represents that these parameters are defined in Ref.\cite{zAcA}.
Note in particular that a minus sign in the relation (\ref{eq:redefine cAL}).
A factor $1/u_0$ in  
Eqs.(\ref{eq:redefine cA-})$-$(\ref{eq:redefine cAR}) 
is due to link variables in the 
point splitting operators of
Eqs.(\ref{Eq:lattDrvtvR}) and (\ref{Eq:lattDrvtvL}).
In Eq.(\ref{eq:redefine cA4+}) we
multiply an extra factor $(\mqlomf+\mQlomf)/{\rm
sinh}(\mqlomf+\mQlomf)$ 
since $c_{A_4}^{+,{\rm PT}}$ in Ref.\cite{zAcA} is a coefficient of 
$(\mqlomf+\mQlomf)\bar{q}(q)Q(p)$ while
$c_{A_4}^{+}$ in Eq.(\ref{eq:RHQcrrntImpr}) is a coefficient 
${\rm sinh}(\mqlomf+\mQlomf)\bar{q}(q)Q(p)$.


\clearpage
\begin{table}
\begin{center}
\caption{
Simulation parameters for the RHQ action 
with the Iwasaki gauge action(upper) and the plaquette gauge action(lower).
\label{Tab:Sim:Iwasaki+RHQ}}
\begin{tabular}{c|ccccccc}
      \hline \hline
   &  \multicolumn{7}{c}{Iwasaki}\\    
   &  \multicolumn{1}{c}{flavor}
   &  \multicolumn{1}{c}{$\kappa$}
   &  \multicolumn{1}{c}{$M_{PS}/M_V$}
   &  \multicolumn{1}{c}{$\nu$}
   &  \multicolumn{1}{c}{$r_s$}
   &  \multicolumn{1}{c}{$c_{B}$}
   &  \multicolumn{1}{c}{$\omega=c_{E}/c_{B}$}
   \\ \hline
   &  $\kappa_1$
   &  $0.13295$
   & $    0.5567(   36)$
   &  $1.0$
   &  $1.0$
   &  $1.50$
   &  $1.0$
   \\ 
      light
   &  $\kappa_2$
   &  $0.13222$
   & $    0.6898(   23)$
   &  $1.0$
   &  $1.0$
   &  $1.50$
   &  $1.0$
   \\ 
   &  $\kappa_3$
   &  $0.13138$
   & $    0.7734(   16)$
   &  $1.0$
   &  $1.0$
   &  $1.50$
   &  $1.0$
   \\ \hline
   &  $\kappa_4$
   &  $0.11513$
   & $    0.9372(    7)$
   &  $1.03160$
   &  $1.12787$
   &  $1.66304$
   &  $0.92064$
   \\ 
      heavy
   &  $\kappa_5$
   &  $0.10524$
   & $    0.9680(    4)$
   &  $1.05935$
   &  $1.20160$
   &  $1.75930$
   &  $0.88889$
   \\ 
   &  $\kappa_6$
   &  $0.09455$
   & $    0.9813(    2)$
   &  $1.10040$
   &  $1.29777$
   &  $1.88490$
   &  $0.85628$
   \\ 
   &  $\kappa_7$
   &  $0.07841$
   & $    0.9901(    1)$
   &  $1.19159$
   &  $1.48857$
   &  $2.13426$
   &  $0.81003$
   \\ \hline \hline
      
   &  \multicolumn{7}{c}{plaquette}\\    
   &  \multicolumn{1}{c}{flavor}
   &  \multicolumn{1}{c}{$\kappa$}  
   &  \multicolumn{1}{c}{$M_{PS}/M_V$}
   &  \multicolumn{1}{c}{$\nu$}
   &  \multicolumn{1}{c}{$r_s$}
   &  \multicolumn{1}{c}{$c_{B}$}
   &  \multicolumn{1}{c}{$\omega=c_{E}/c_{B}$}
   \\ \hline
   &  $\kappa_1$
   &  $0.13449$
   &  $0.5492(  173)$
   &  $1.0$
   &  $1.0$
   &  $1.769$
   &  $1.0$
   \\ 
      light
   &  $\kappa_2$
   &  $0.13373$
   &  $0.7088(   27)$
   &  $1.0$
   &  $1.0$
   &  $1.769$
   &  $1.0$
   \\ 
   &  $\kappa_3$
   &  $0.13298$
   &  $0.7837(   16)$
   &  $1.0$
   &  $1.0$
   &  $1.769$
   &  $1.0$
   \\ \hline
   &  $\kappa_4$
   &  $0.11456$
   &  $0.9345(    5)$
   &  $1.04161$
   &  $1.16034$
   &  $2.02423$
   &  $0.91709$
   \\ 
      heavy
   &  $\kappa_5$
   &  $0.10190$
   &  $0.9728(    2)$
   &  $1.08301$
   &  $1.26201$
   &  $2.17790$
   &  $0.88345$
   \\ 
   &  $\kappa_6$
   &  $0.09495$
   &  $0.9808(    2)$
   &  $1.11284$
   &  $1.32840$
   &  $2.27776$
   &  $0.86586$
   \\ 
   &  $\kappa_7$
   &  $0.07490$
   &  $0.9911(    1)$
   &  $1.23871$
   &  $1.58259$
   &  $2.66050$
   &  $0.81742$
   \\ \hline \hline
\end{tabular}
\end{center}
\end{table}
\begin{table}
\begin{center}
\caption{Simulation parameters for the heavy clover quark action 
with the Iwasaki gauge action(upper) and the plaquette gauge action(lower).
\label{Tab:Sim:Iwasaki+CL}}
\begin{tabular}{c|ccccccc}
      \hline \hline
      
   &  \multicolumn{7}{c}{Iwasaki}\\    
   &  \multicolumn{1}{c}{flavor}
   &  \multicolumn{1}{c}{$\kappa$}  
   &  \multicolumn{1}{c}{$M_{PS}/M_V$}
   &  \multicolumn{1}{c}{$\nu$}
   &  \multicolumn{1}{c}{$r_s$}
   &  \multicolumn{1}{c}{$c_{B}$}
   &  \multicolumn{1}{c}{$\omega=c_{E}/c_{B}$}
   \\ \hline
      light
   &  $\kappa_3$
   &  $0.13138$
   &  $0.7734(   16)$
   &  $1.0$
   &  $1.0$
   &  $1.50$
   &  $1.0$
   \\ \hline
   &  $\tilde{\kappa}_4$
   &  $0.1256$
   &  $0.9353(    5)$
   &  $1.0$
   &  $1.0$
   &  $1.50$
   &  $1.0$
   \\ 
      heavy
   &  $\tilde{\kappa}_5$
   &  $0.1186$
   &  $0.9717(    3)$
   &  $1.0$
   &  $1.0$
   &  $1.50$
   &  $1.0$
   \\ 
   &  $\tilde{\kappa}_6$
   &  $0.1119$
   &  $0.9836(    2)$
   &  $1.0$
   &  $1.0$
   &  $1.50$
   &  $1.0$
   \\ \hline \hline
      
   &  \multicolumn{7}{c}{plaquette}\\    
   &  \multicolumn{1}{c}{flavor}
   &  \multicolumn{1}{c}{$\kappa$}  
   &  \multicolumn{1}{c}{$M_{PS}/M_V$}
   &  \multicolumn{1}{c}{$\nu$}
   &  \multicolumn{1}{c}{$r_s$}
   &  \multicolumn{1}{c}{$c_{B}$}
   &  \multicolumn{1}{c}{$\omega=c_{E}/c_{B}$}
   \\ \hline
      light
   &  $\kappa_3$
   &  $0.13298$
   &  $0.7837(   16)$
   &  $1.0$
   &  $1.0$
   &  $1.769$
   &  $1.0$
   \\ \hline
   &  $\tilde{\kappa}_4$
   &  $0.12780$
   &  $0.9359(   8)$
   &  $1.0$
   &  $1.0$
   &  $1.769$
   &  $1.0$
   \\ 
      heavy
   &  $\tilde{\kappa}_5$
   &  $0.11900$
   &  $0.9770(   3)$
   &  $1.0$
   &  $1.0$
   &  $1.769$
   &  $1.0$
   \\ 
   &  $\tilde{\kappa}_6$
   &  $0.11480$
   &  $0.9834(   2)$
   &  $1.0$
   &  $1.0$
   &  $1.769$
   &  $1.0$
   \\ \hline \hline
\end{tabular}
\end{center}
\end{table}
\begin{table}
\begin{center}
\caption{Smearing parameters.
\label{Tab:Sim:Smear para}}
\begin{tabular}{c|ccc|ccc|ccc|ccc}
      \hline \hline
      action
   & \multicolumn{3}{c}{Iwasaki+RHQ} \vline
   & \multicolumn{3}{c}{Iwasaki+CL} \vline    
   & \multicolumn{3}{c}{plaquette+RHQ} \vline 
   & \multicolumn{3}{c}{plaquette+CL}    \\
      
   &  flavor
   &  $A$
   &  $B$
   &  flavor
   &  $A$
   &  $B$
   &  flavor
   &  $A$
   &  $B$
   &  flavor
   &  $A$
   &  $B$
   \\ \hline
   &  $\kappa_1$
   &  $1.28$
   &  $0.28$
   &
   &
   &
   &  $\kappa_1$
   &  $1.28$
   &  $0.35$
   &
   &
   &
   \\ 
      light
   &  $\kappa_2$
   &  $1.25$
   &  $0.3$
   &
   &
   &
   &  $\kappa_2$
   &  $1.28$
   &  $0.35$
   &
   &
   &
   \\ 
   &  $\kappa_3$
   &  $1.25$
   &  $0.32$
   &  $\kappa_3$
   &  $1.25$
   &  $0.32$
   &  $\kappa_3$
   &  $1.28$
   &  $0.35$
   &  $\kappa_3$
   &  $1.28$
   &  $0.35$
   \\ \hline
   &  $\kappa_4$
   &  $1.25$
   &  $0.50$
   &  $\tilde{\kappa}_4$
   &  $1.25$
   &  $0.50$
   &  $\kappa_4$
   &  $1.25$
   &  $0.50$
   &  $\tilde{\kappa}_4$
   &  $1.25$
   &  $0.50$
   \\ 
      heavy
   &  $\kappa_5$
   &  $1.25$
   &  $0.58$
   &  $\tilde{\kappa}_5$
   &  $1.25$
   &  $0.65$
   &  $\kappa_5$
   &  $1.25$
   &  $0.58$
   &  $\tilde{\kappa}_5$
   &  $1.25$
   &  $0.65$
   \\ 
   &  $\kappa_6$
   &  $1.25$
   &  $0.65$
   &  $\tilde{\kappa}_6$
   &  $1.25$
   &  $0.82$
   &  $\kappa_6$
   &  $1.25$
   &  $0.60$
   &  $\tilde{\kappa}_6$
   &  $1.25$
   &  $0.65$
   \\ 
   &  $\kappa_7$
   &  $1.25$
   &  $1.00$
   &
   &
   &
   &  $\kappa_7$
   &  $1.25$
   &  $0.8$
   &
   &
   &
   \\ \hline \hline
\end{tabular}
\end{center}
\end{table}
\begin{table}
\begin{center}
\caption{Fitting range from $t_{\rm min}$ to $t_{\rm max}$ for the two-point
functions.   
\label{tab:fitrng}}
\begin{tabular}{cccccc}
\hline
\hline
&&& Iwasaki & plaquette \\
correlator & system & source & $t_{\rm min}/t_{\rm
max}$ & $t_{\rm min}/t_{\rm max}$ \\ 
\hline
$\langle V V^{\dagger}\rangle$ & H-L & S & 10/22 & 7/17 \\
$\langle P P^{\dagger}\rangle$ & H-L & S & 10/22 & 9/22 \\
$\langle P P^{\dagger}\rangle$ & H-L & P & 13/22 & 14/22 \\
$\langle P P^{\dagger}\rangle$ with $|{\vec p}|\ne 0$ & H-L & S & 8/20 & 9/22
\\
\hline
$\langle V V^{\dagger}\rangle$ & H-H & S & 12/22 & 11/21 \\
$\langle P P^{\dagger}\rangle$ & H-H & S & 12/22 & 11/21 \\
$\langle P P^{\dagger}\rangle$ & H-H & P & 16/22 & 18/23 \\
$\langle P P^{\dagger}\rangle$ with $|{\vec p}|\ne 0$ & H-H & S & 12/20 &
10/22 \\
\hline
\end{tabular}
\end{center}
\end{table}
\begin{table}
\begin{center}
\caption{
Fitting results for chiral extrapolation of $M_{PS}^2$
for the light-light pseudoscalar meson.
\label{Tab:Fit:Mpi}}
\begin{tabular}{c|cccc}
      \hline \hline
      action
   &  \multicolumn{1}{c}{$A$}
   &  \multicolumn{1}{c}{$\kappa_{crit}$}
   &  \multicolumn{1}{c}{$\chi^2$/dof}
   \\ \hline
      Iwasaki+RHQ
   &  $0.7583( 15)$
   &  $0.133802(4)$
   &  $6.3$
   \\ 
      plaquette+RHQ
   &  $0.7886( 38)$
   &  $0.135247(9)$
   &  $0.7$
   \\ \hline \hline
\end{tabular}
\end{center}
\end{table}
\begin{table}
\begin{center}
\caption{Hopping parameters at the physical points.
\label{tab:PhysP}}
\begin{tabular}{c|cccccc}
      \hline \hline
      action
   &  \multicolumn{1}{c}{$\kappa_{ud}$} 
   &  \multicolumn{1}{c}{$\kappa_{s}(K)$}
   &  \multicolumn{1}{c}{$\kappa_c(J/\psi, M_{\rm
      pole})$}
   &  \multicolumn{1}{c}{$\kappa_c(J/\psi, M_{\rm
      kin})$}
   &  \multicolumn{1}{c}{$\kappa_c(D_s, M_{\rm
      pole})$}
   &  \multicolumn{1}{c}{$\kappa_c(D_s, M_{\rm
      kin})$}
   \\ \hline
      Iwasaki+RHQ
   &  $0.133749(4)$
   &  $0.132422(4)$
   &  $0.099414(22)$
   &  $0.102362(377)$
   &  $0.099640(49)$
   &  $0.101528(862)$
   \\ 
      plaquette+RHQ
   &  $0.135200(9)$
   &  $0.134026(5)$
   &  $0.100593(21)$
   &  $0.102610(343)$
   &  $0.100669(35)$
   &  $0.102402(810)$
   \\ \hline \hline
\end{tabular}
\end{center}
\end{table}
\begin{table}
\begin{center}
\caption{
Fitting results for the heavy-heavy hyperfine splitting
as a function of the vector meson mass.
\label{Tab:Fit:Mcc}}
\begin{tabular}{c|cccc|cccc}
      \hline \hline
   &  \multicolumn{4}{c}{$M_{\rm pole}$} \vline
   &  \multicolumn{4}{c}{$M_{\rm kin}$}
   \\ 
      action  
   &  \multicolumn{1}{c}{$A$}
   &  \multicolumn{1}{c}{$B$}
   &  \multicolumn{1}{c}{$C$}
   &  \multicolumn{1}{c}{$\chi^2$/dof} \vline
   &  \multicolumn{1}{c}{$A$}
   &  \multicolumn{1}{c}{$B$}
   &  \multicolumn{1}{c}{$C$}
   &  \multicolumn{1}{c}{$\chi^2$/dof}
    \\ \hline
      Iwasaki+RHQ
   &  $0.0452(  14)$
   &  $0.0156(  43)$
   &  $0.0009(  26)$
   &  $0.0018$
   &  $0.0676(  23)$
   &  $-0.0120(  78)$
   &  $0.0122(  50)$
   &  $0.001$
   \\ 
      plaquette+RHQ
   &  $0.0524(  14)$
   &  $-0.0028(  33)$
   &  $0.0081(  19)$
   &  $0.03$
   &  $0.0745(  20)$
   &  $-0.0403(  59)$
   &  $0.0288(  34)$
   &  $0.10$
   \\ \hline \hline
\end{tabular}
\end{center}
\end{table}
\begin{table}
\begin{center}
\caption{Charmonium and charmed-strange meson hyper-fine 
splittings in unit of GeV with $M_{\rm pole}$ and
$M_{\rm kin}$. The lattice spacing is determined 
by the Sommer scale $r_0=0.5$fm.
\label{Tab:HFS:CC}}
\begin{tabular}{c|ccccc}
   \hline \hline
      
   &  \multicolumn{2}{c}{Iwasaki}
   &  \multicolumn{2}{c}{plaquette} & \\
   &  \multicolumn{1}{c}{$M_{\rm pole}$}
   &  \multicolumn{1}{c}{$M_{\rm kin}$}
   &  \multicolumn{1}{c}{$M_{\rm pole}$}
   &  \multicolumn{1}{c}{$M_{\rm kin}$}
   &  \multicolumn{1}{c}{expt.} 
\\ \hline
$\Delta M(J/\psi-\eta_c)$ 
   & $    0.0728(    8)$
   & $    0.0847(   20)$
   & $    0.0788(    6)$
   & $    0.0877(   18)$
   & $0.1173(12)$ \\
$\Delta M(D_s^*-D_s)$ 
   & $    0.1243(   28)$
   & $    0.1348(   65)$
   & $    0.1261(   16)$
   & $    0.1358(   57)$
   & $0.1438(4)$
\\ \hline \hline
\end{tabular}
\end{center}
\end{table}
\begin{table}
\begin{center}
\caption{
Fitting results for the heavy-light hyperfine splitting
as a function of the pseudoscalar meson mass.
\label{Tab:Fit:HFS:CS}}
\begin{tabular}{c|cccc|cccc}
      \hline \hline
   &  \multicolumn{4}{c}{$M_{\rm pole}$} \vline
   &  \multicolumn{4}{c}{$M_{\rm kin}$} \\
      action
   &  \multicolumn{1}{c}{$A$}
   &  \multicolumn{1}{c}{$B$}
   &  \multicolumn{1}{c}{$C$}
   &  \multicolumn{1}{c}{$\chi^2$/dof} \vline
   &  \multicolumn{1}{c}{$A$}
   &  \multicolumn{1}{c}{$B$}
   &  \multicolumn{1}{c}{$C$}
   &  \multicolumn{1}{c}{$\chi^2$/dof} 
   \\ \hline
      Iwasaki+RHQ
   &  $0.0602(  35)$
   &  $0.0028(  17)$
   &  $-0.0628( 167)$
   &  $0.13$
   &  $0.0649(  75)$
   &  $0.0012(  41)$
   &  $-0.0106(  337)$
   &  $0.03$
   \\ 
      plaquette+RHQ
   &  $0.0545(  20)$
   &  $0.0022(  10)$
   &  $-0.0523( 121)$
   &  $0.22$
   &  $0.0549(  70)$
   &  $0.0049(  36)$
   &  $-0.0224(  497)$
   &  $0.61$
   \\ \hline \hline
\end{tabular}
\end{center}
\end{table}
\begin{table}
\begin{center}
\caption{
Fitting results for the heavy-light pseudoscalar meson decay
constant determined from $A_4$ as a function of the
pseudoscalar meson mass.
\label{Tab:Fit:fcsA4}}
\begin{tabular}{c|ccccc}
      \hline \hline
   &  \multicolumn{5}{c}{$M_{\rm pole}$}
   \\ 
      action
   &  \multicolumn{1}{c}{$A$}
   &  \multicolumn{1}{c}{$B$}
   &  \multicolumn{1}{c}{$C$}
   &  \multicolumn{1}{c}{$D$}
   &  \multicolumn{1}{c}{$\chi^2$/dof}
   \\ \hline
      Iwasaki+RHQ
   & $    0.2204(   93)$
   & $   -0.1337(  113)$
   & $    0.0312(   39)$
   & $    0.1780(  199)$
   & $    0.43$
   \\ 
      plaquette+RHQ
   & $    0.1285(   20)$
   & $   -0.0217(   26)$
   & $   -0.0065(   10)$
   & $    0.1967(  140)$
   & $    2.9$ \\
      \hline \hline
   &  \multicolumn{5}{c}{$M_{\rm kin}$}
   \\ 
      action
   &  \multicolumn{1}{c}{$A$}
   &  \multicolumn{1}{c}{$B$}
   &  \multicolumn{1}{c}{$C$}
   &  \multicolumn{1}{c}{$D$}
   &  \multicolumn{1}{c}{$\chi^2$/dof}
   \\ \hline
      Iwasaki+RHQ
   & $    0.2275(   96)$
   & $   -0.1470(  127)$
   & $    0.0368(   47)$
   & $    0.1820(  210)$
   & $    0.33$
   \\ 
      plaquette+RHQ
   & $    0.1326(   29)$
   & $   -0.0241(   49)$
   & $   -0.0071(   22)$
   & $    0.1903(  131)$
   & $    1.7$
   \\ \hline \hline
\end{tabular}
\end{center}
\end{table}
\begin{table}
\begin{center}
\caption{
Fitting results for the heavy-light pseudoscalar meson decay
constant determined from $A_k$ as a function of the
pseudoscalar meson mass.
\label{Tab:Fit:fcsAk}}
\begin{tabular}{c|ccccc}
      \hline \hline
   &  \multicolumn{5}{c}{$M_{\rm pole}$}
   \\ 
      action
   &  \multicolumn{1}{c}{$A$}
   &  \multicolumn{1}{c}{$B$}
   &  \multicolumn{1}{c}{$C$}
   &  \multicolumn{1}{c}{$D$}
   &  \multicolumn{1}{c}{$\chi^2$/dof}
   \\ \hline
      Iwasaki+RHQ
   & $    0.1892(   88)$
   & $   -0.0955(  110)$
   & $    0.0174(   39)$
   & $    0.1788(  190)$
   & $    0.44$
   \\ 
      plaquette+RHQ
   & $    0.1709(   57)$
   & $   -0.0811(   73)$
   & $    0.0144(   27)$
   & $    0.1338(  306)$
   & $    0.55$
   \\ \hline \hline
   &  \multicolumn{5}{c}{$M_{\rm kin}$}
   \\ 
      action
   &  \multicolumn{1}{c}{$A$}
   &  \multicolumn{1}{c}{$B$}
   &  \multicolumn{1}{c}{$C$}
   &  \multicolumn{1}{c}{$D$}
   &  \multicolumn{1}{c}{$\chi^2$/dof}
   \\ \hline
      Iwasaki+RHQ
   & $    0.1975(   96)$
   & $   -0.1104(  129)$
   & $    0.0235(   50)$
   & $    0.1836(  198)$
   & $    0.29$
   \\ 
      plaquette+RHQ
   & $    0.1746(   54)$
   & $   -0.0868(   70)$
   & $    0.0163(   28)$
   & $    0.1370(  311)$
   & $    0.86$
   \\ \hline \hline
\end{tabular}
\end{center}
\end{table}
\begin{table}
\begin{center}
\caption{$D_s$ meson decay constants in unit of GeV
determined from $A_k$ and $A_4$
using $M_{\rm pole}$ as well as $M_{\rm kin}$.  
The lattice spacing is determined by the Sommer scale
$r_0=0.5$fm. The experimental value for $f_{D_s}$ is take 
from Ref.\cite{fDs_expt}.
\label{Tab:fCS}}
\begin{tabular}{c|ccccc}
   \hline \hline
   &  \multicolumn{2}{c}{Iwasaki}
   &  \multicolumn{2}{c}{plaquette} & \\
   &  \multicolumn{1}{c}{$M_{\rm pole}$}
   &  \multicolumn{1}{c}{$M_{\rm kin}$}
   &  \multicolumn{1}{c}{$M_{\rm pole}$}
   &  \multicolumn{1}{c}{$M_{\rm kin}$}
   &  \multicolumn{1}{c}{expt.} 
\\ \hline
$f_{D_s}(A_4)$ 
   & $    0.2506(   49)$
   & $    0.2496(   48)$
   & $    0.2291(   22)$
   & $    0.2304(   24)$
   & $    0.282(16)(7)  $
\\ 
$f_{D_s}(A_k)$
   & $    0.2373(   47)$
   & $    0.2369(   46)$
   & $    0.2305(   31)$
   & $    0.2304(   30)$
   & $    0.282(16)(7)    $
\\ \hline \hline
\end{tabular}
\end{center}
\end{table}
\begin{table}
\begin{center}
\caption{
The final results of the charmonium hyperfine mass splitting,
the charmed-strange meson hyperfine splitting and the
$D_s$ meson decay constant in unit of GeV.
The first error is statistical and the second and the third ones
are the cutoff errors explained in the text.
The lattice spacing is determined by the Sommer scale $r_0=0.5$fm.
\label{Tab:HFS:CC:Fnl}}
\begin{tabular}{c|ccccccc}
   \hline \hline
      
   &  \multicolumn{1}{c}{Iwasaki}
   &  \multicolumn{1}{c}{plaquette}
   &  \multicolumn{1}{c}{expt.} 
\\ \hline
$\Delta M(J/\Psi-\eta_c)$ 
   & $    0.0728(    8)(_{-0}^{+119})$
   & $    0.0788(    6)(_{-0}^{+89})$
   & $0.1173(12)$ \\
$\Delta M(D_s^*-D_s)$ 
   & $    0.1243(   28)(_{-0}^{+105})$
   & $    0.1261(   16)(_{-0}^{+97})$
   & $0.1438(4)$ \\
$f_{D_s}(A_4)$ 
   & $    0.2506(   49)(_{-0}^{+10})(_{-0}^{+133})$
   & $    0.2291(   22)(_{-0}^{+13})(_{-0}^{+14})$
   & $0.267(33)$
\\ \hline \hline
\end{tabular}
\end{center}
\end{table}

\clearpage


\begin{thebibliography}{9999999999999}


\bibitem{akt} S.~Aoki, Y.~Kuramashi and S.~Tominaga,
Prog.~Theor.~Phys.~{\bf 109} (2003) 383.

\bibitem{sym1} K.~Symanzik, in 
{\it Mathematical Problems In Theoretical Physics}, 
eds. R.~Schrader {\it et al.}, 
Lecture Notes in Physics, Vol. 153 (Springer, New York,
1982).

\bibitem{sym2} K.~Symanzik, 
Nucl. Phys. {\bf B226} (1983), 187; 205. 

\bibitem{sw} B.~Sheikholeslami and R.~Wohlert, 
Nucl. Phys. {\bf B259} (1985), 572. 

\bibitem{alpha} M.~L\"{u}scher, S.~Sint, R.~Sommer and
P.~Weisz Nucl. Phys. {\bf B478} (1996), 365. 

\bibitem{onshell} M.~L\"{u}scher and P.~Weisz, 
Commun. Math. Phys. {\bf 97} (1985), 59. 

\bibitem{Christ}  N. H.~Christ, M.~Li, H.-W.~Lin, hep-lat/0608006;
H.-W.~Lin, and N. H.~Christ, hep-lat/0608005.

\bibitem{cEcB} S.~Aoki, Y.~Kayaba and Y.~Kuramashi, 
Nucl.~Phys.~{\bf B}~{\bf 697} (2004) 271.

\bibitem{zAcA} S.~Aoki, Y.~Kayaba and Y.~Kuramashi,
Nucl.~Phys.~{\bf B}~{\bf 689} (2004) 127.

\bibitem{KrmshNote} S.~Aoki, Y.~Kayaba, Y.~Kuramashi and
N.~Yamada, 
hep-lat/0409001, to appear in
Progress of Theoretical Physics Supplement.

\bibitem{iwasaki}
Y.~Iwasaki,
preprint, UTHEP-118 (Dec. 1983), unpublished.

\bibitem{r0Necco} S.~Necco,
Nucl.~Phys.~{\bf B}~{\bf 683} (2004) 137.

\bibitem{r0Sommer} M.~Guagnelli, R.~Sommer, H.~Wittig,
Nucl.~Phys.~{\bf B}~{\bf 535} (1998) 389.

\bibitem{NPCL.CPPACS} CP-PACS and JLQCD Collaborations:
S.~Aoki {\it et al.}, Phys.~Rev.~{\bf D73} (2006) 034501.

\bibitem{NPCL.ALPHA} M.~L{\"u}scher, S.~Sint, R.~Sommer,
P.~Weisz and U.~Wolf, Nucl.~Phys.~{\bf B}~{\bf 491}
(1997) 323.

\bibitem{NPzAcA:LALM:3}
T.~Bhattacharya, R. Gupta, W. Lee and S. Sharpe,
Phys. Rev. D 62 (2001) 074505.

\bibitem{gmass} 
S.~Aoki, K.~Nagai, Y.~Taniguchi and A.~Ukawa,
Phys.~Rev.~{\bf D58} (1998) 074505; 
Y.~Taniguchi and A.~Ukawa,
Phys.~Rev.~{\bf D58} (1998) 114503.


\bibitem{fDs_expt}
CLEO Collaboration, M.~Artuso {\it et al.}, hep-ex/0607074;
Phys. Rev. Lett. 95 (2005) 251801.

\bibitem{Aniso:CC:Spctrm} 
CP-PACS Collaboration: M~.Okamoto {\it et al.}, 
Phys. Rev. {\bf D65} (2002) 094508.

\bibitem{fDs.UKQCD}
UKQCD Collaboration: K.~C.~Bowler {\it et al.},
Nucl. Phys. B619 (2001) 507.

\bibitem{fDs.ALPHA} ALPHA Collaboration: A.~J{\" u}ttner
and J.~Rolf, Phys. Lett. B560 (2003) 59.

\bibitem{latt05} 
CP-PACS Collaboration: Y.~Kuramashi {\it et al.}, 
Proc.~Sci.~LAT2005 (2005) 226. 

\bibitem{Tadpole.val}
S. Aoki, T. Izubuchi, Y. Kuramashi and Y. Taniguchi,
Phys. Rev. D67 (2003) 094502.

\bibitem{CP-PACS:fHL:NRQCD:Nf=2}
CP-PACS Collaboration: A.~Ali.~Khan {\it et al.},
Phys.Rev.~{\bf D64} (2001) 054504.



\end{thebibliography}
\end{document}